\documentclass[fleqn,usenatbib]{mnras}

\usepackage{newtxtext,newtxmath}

\usepackage{soul}
\usepackage[T1]{fontenc}

\DeclareRobustCommand{\VAN}[3]{#2}
\let\VANthebibliography\thebibliography
\def\thebibliography{\DeclareRobustCommand{\VAN}[3]{##3}\VANthebibliography}

\usepackage{graphicx}	
\usepackage{amsmath}	
\usepackage{lscape}

\title[Thermonuclear bursts from MAXI J1816--195]
{The study of thermonuclear X-ray bursts in accreting millisecond pulsar MAXI J1816--195 with {\it NuSTAR} and {\it NICER}}

\author[M. Mandal et al.]{
Manoj Mandal,$^{1}$\thanks{E-mail: manojmandal@mcconline.org.in}
Sabyasachi Pal,$^{1}$\thanks{E-mail: sabya.pal@gmail.com}
Jaiverdhan Chauhan,$^{2}$\thanks{E-mail: jaiverdhan.chauhan@montana.edu}
Anne Lohfink,$^{2}$
Priya Bharali$^{3}$\\
$^{1}$Midnapore City College, Kuturia, Bhadutala, West Bengal, 721129, India \\
$^{2}$Department of Physics, Montana State University, P.O. Box 173840, Bozeman, MT 59717-3840, USA \\
$^{3}$Mahatma Gandhi Government Arts College, Puducherry, India\\
\\
}

\date{Accepted XXX. Received YYY; in original form ZZZ}

\pubyear{2015}

\begin{document}
\label{firstpage}
\pagerange{\pageref{firstpage}--\pageref{lastpage}}
\maketitle

\begin{abstract}
The millisecond pulsar MAXI J1816--195 was recently discovered in an outburst by {\it MAXI} in 2022 May. We study different properties of the pulsar using data from {\it NuSTAR} and {\it NICER} observations. 
The unstable burning of accreted material on the surface of neutron stars induces thermonuclear (Type-I) bursts. Several such thermonuclear bursts have been detected by MAXI J1816--195 during its outburst. We investigate the evolution of the burst profile with flux and energy using {\it NuSTAR} and {\it NICER} observations. During the {\it NuSTAR} observation, a total of four bursts were detected from the source. The duration of each burst is around $\sim$ 30\,s and the ratio of peak to persistent count rate is $\sim$ 26 as seen from the {\it NuSTAR} data. The burst profiles are modelled using a sharp linear rise and exponential decay function to determine the burst timing parameters. The burst profiles show a relatively long tail at lower energies. The broadband time-resolved spectra during the burst periods are successfully modelled with a combination of an absorbed blackbody along with a non-thermal component to account for the persistent emission. 
From our modelling results, we are able to estimate the maximum apparent emitting area of the blackbody of the neutron star to be $\sim$12.5\,km during the peak of the outburst and the maximum distance to the object to be $8.7$\,kpc. Our findings for the mass accretion rate and the alpha factor indicate the stable burning of hydrogen via a hot CNO cycle during the bursts.
\end{abstract}

\begin{keywords}
accretion, accretion discs -- stars: neutron – X-ray: binaries -- X-rays: bursts – X-rays: individual: MAXI J1816--195
\end{keywords}


\section{Introduction}
\label{intro}
MAXI J1816--195, an accreting X-ray pulsar, was discovered by the Monitor of All-sky X-ray Image\footnote{\url{http://maxi.riken.jp/top/index.html}} \citep[{\it MAXI};][]{Ma09} on 2022 June 7 \citep{Ne22}, and has subsequently been followed-up by different multi-wavelength observations \citep{Be22, Br22, Ke22}. Importantly, \citet{Bu22} found pulsations at 528.6\,Hz from MAXI J1816--195 using {\it NICER}, confirming the nature of the source as an accreting millisecond X-ray pulsar. The timing results revealed that MAXI J1816--195 possesses an orbital period of 4.8\,hours \citep{Bul22}. \textit{NICER} observations also found 15 thermonuclear bursts from the source during the span of its outburst \citep{Bul22}. The source was also observed by {\it NuSTAR} and the pulsation at 528\,Hz was confirmed, and an X-ray reflection component was found \citep{Ch22}.  MAXI J1816--195 was also studied using {\it Insight}-HXMT and 73 bursts from the source have been detected \citep{Pe22}. The broadband persistent spectra were modelled using an absorbed convolution thermal Comptonization component with input seed photons from the accretion disk. During  bursts, an absorbed blackbody model is used to fit the {\it HXMT} spectra and $f_a$ model is applied to account for the variable persistent emission.

The first thermonuclear (Type-I) X-ray burst in a pulsar was discovered more than 40 years ago in 1976 \citep{Gr76}; it provided a new arena for the study of different parameters of neutron stars \citep[such as mass, radius, spin frequency, photospheric radius expansion;][]{Ba10}. During a Type-I burst, the surface emission is typically more than 10 times brighter than the remaining X-ray emission. As a result, it may be possible to isolate the surface radiation from the total emission and utilize it to accurately determine the neutron star parameters \citep{Ba10}. An X-ray burst is powered by a low-mass companion star from which the neutron star (NS) accretes gas through an accretion disc. Matter gathers on the stellar surface of the neutron stars from the low-mass binary companion. During the burst, thermonuclear burning transforms the accreted hydrogen and helium on the surface of the NS into heavier elements \citep{Le93, St03, Sc06}. The burning layer will ignite completely on a time scale of a few seconds and cause a burst if radiative cooling is slower than the energy generation rate.

 The rapid increase in the X-ray intensity during a thermonuclear burst is mainly due to the unstable ignition of fuel on the surface of the neutron star \citep{Ga08}, which returns to the pre-burst level after a short time by following an exponential decay \citep{Le93, Ga08}. 
 During the burst, the rapid flux rise happens in a short period of  0.5--5 s and the decay happens over  10--100 s \citep{Le93, Ba10}. 
To date, Type-I bursts have been found in more than 100 sources \citep{Li07}. Not all bursts have the same burst profile, both single and multi-peaked burst profiles have been reported depending on the source. Single-peaked bursts are typically observed in most of the thermonuclear bursts but double-peaked bursts are observed for several sources like GX17+2 \citep{Ku02}, 4U 1709--267 \citep{Jo04}, 4U 1636--53 \citep{Wa07}, and 4U 1608--52 \citep{Pe89}, MXB 1730--335 \citep{Ba14}. Triple-peaked bursts are relatively rare to observe, and observed in 4U 1636--53 \citep{va86, Zh09}. \citet{Li21} found several multi-peaked Type-I bursts from 4U 1636--536 using {\it RXTE}, and an extraordinary quadruple-peaked burst was also observed for the same source.

At higher energies, double-peaked profiles were observed for several sources due to photospheric expansion when the peak flux reached the Eddington luminosity \citep{Le93}. Near a constant luminosity ($L$), the photospheric radius ($R$) increases and the effective temperature decreases as, $T_{\textrm{eff}} = \left(\frac{L}{4\pi\sigma}\right)^{1/4} R^{-1/2}$ \citep{Le93}. The time-resolved continuum spectra are modelled with an absorbed blackbody component by assuming the entire surface of the neutron star emits like a blackbody \citep{Va78, Ku03}. The Eddington limit may be reached at the peak of the burst, which would result in extremely high radiation pressure and allow the expansion of the photosphere of the neutron star \citep{Ta84}. The blackbody radius continues to increase during the rising phase of a burst, reaching a maximum value. The photospheric radius started to decrease during the decay phase of the burst. This process continues until the photospheric radius reaches its original value and the neutron star surface starts cooling. At the end of the process, the photospheric radius becomes equal to the radius of the neutron star, and at the touchdown phase, the temperature has the highest value along with a lower blackbody radius \citep{Le93, Ku03}. The radius of a neutron star can be measured using  time-resolved spectroscopy during the thermonuclear burst \citep[during the cooling phase after the touchdown;][]{Le93, Va78, Ga08}. The blackbody temperature shows two maxima in the temperature profile during the Type-I burst, the second maxima implies the touchdown phase \citep{Ga08}. Previously, Type-I X-ray bursts were observed from GRS 1741.9--2853 during the outburst of May 2020 using {\it NuSTAR} which revealed multi-peaked bursts and a photospheric radius expansion (PRE) during the peak of the burst \citep{Pi21}.

Previous work has shown that the count rate, blackbody temperature, and apparent emitting area of the blackbody showed significant variation during the X-ray bursts in several sources. It is this variability that can be used to estimate the evolution of the photosphere radius of the neutron star. The time-resolved spectroscopy of GX 3+1 using {\it Astrosat} observations suggested that there was a photospheric radius expansion for this source during the rising phase of the burst and, using the burst parameters, the source distance was estimated to be $\sim$ 9.3 kpc \citep{Na22}. During the decay phase of an X-ray burst, the photospheric radius decreases and reaches a value of $\sim$6.81 km. Time-resolved spectroscopy was performed for 4U 1636--536 using {\it Astrosat} data, which indicated an expansion of the photospheric radius during an X-ray burst \citep{Be19}. \citet{De21} studied time-resolved spectroscopy for the source Cyg X-2 using the {\it Astrosat} observations, which suggested that the blackbody flux and radius increased in the rising phase of the X-ray burst and that at the same time, the blackbody temperature dropped. The blackbody radius and flux decreased in the decay phase of the burst, and the blackbody temperature increased during the decay of the X-ray burst. Thermonuclear bursts can therefore be used to probe several properties of a source during the burst, including the temperature profile, apparent emitting area of the blackbody, photosphere expansion, and emission mechanism.

 We observed MAXI J1816--195 during the May 2022 outburst using {\it NuSTAR}, which has a good sensitivity up to $\sim$79 keV. The detailed timing and spectral study of MAXI J1816--195 is used to investigate crucial information about the source by using simultaneous {\it NICER} and {\it NuSTAR} observations.
In this paper, we study the timing and spectral properties of MAXI J1816--195 using data from {\it NuSTAR} and {\it NICER} observatories. We have analyzed each burst spectra separately instead of stacking, as the burst profiles seem to be slightly different from each other. We have analyzed two broadband pre and post-burst spectra which are simultaneous with {\it NuSTAR} and {\it NICER}. During the thermonuclear burst, time-resolved spectroscopy is performed using {\it NICER}. This paper is organized as follows: The observational data and reduction procedure are described in Section \ref{obs}. Section \ref{res} summarises the results of spectral and timing analysis. In Section \ref{dis}, we discuss the results obtained. The findings of the study are summarised in Section \ref{con}.

 \begin{figure}
\centering{
\includegraphics[width=9.0 cm]{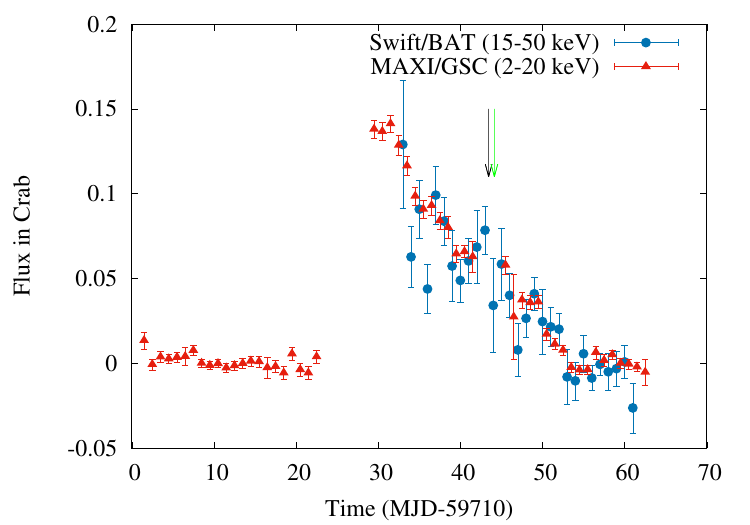}
\caption{An outburst is detected from MAXI J1816--195 using {\it Swift}/BAT and {\it MAXI}/GSC during 2022 May--June. The black arrow indicates the time of the {\it NuSTAR} observation, and the green arrow represents the simultaneous {\it NICER} observation time, respectively. {\it NuSTAR} observed the source during the decay phase of the outburst.
}
\label{fig:BAT}}
\end{figure}

 \begin{figure*}
\centering{
\includegraphics[width=8.5 cm]{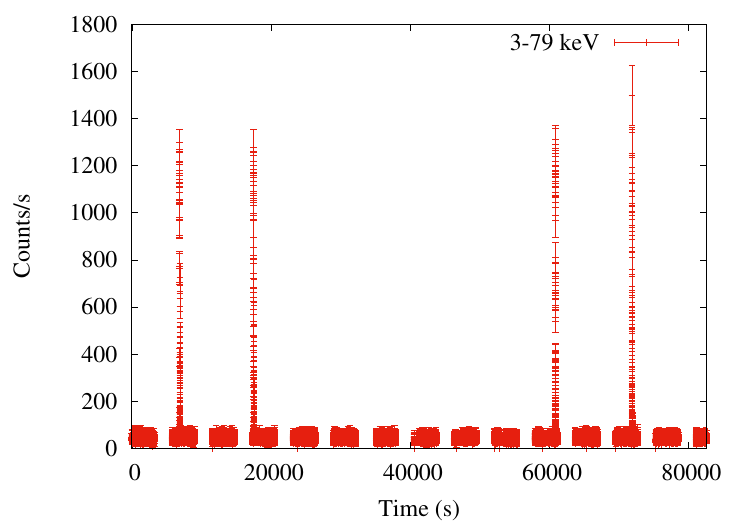}
\includegraphics[width=8.5 cm]{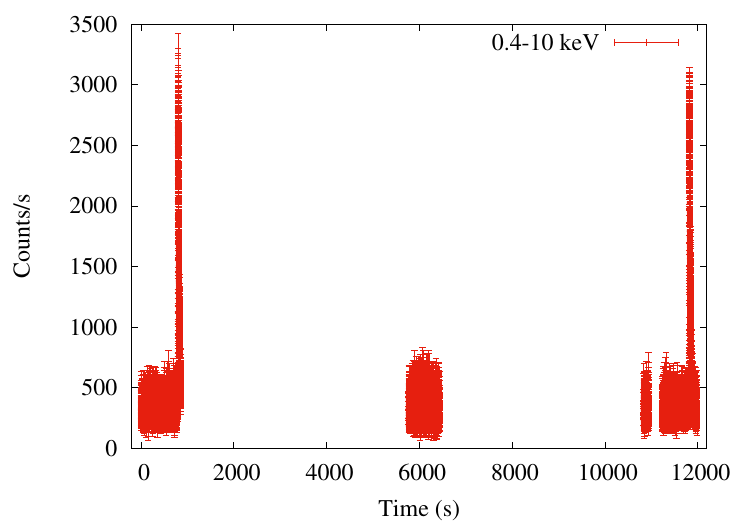}
\caption{The left-hand side panel shows the {\it NuSTAR}-FPMA 1 s binned background corrected light curve in the energy range of 3--79 keV, showing four thermonuclear bursts. The right-hand side panel represents the {\it NICER} (0.4--10 keV) 1 s binned light curve (Obs. ID--533011601) during the simultaneous observation with {\it NuSTAR}. The {\it NICER} light curve shows two bursts, which are also detected with {\it NuSTAR}. The third and fourth bursts in the {\it NuSTAR} light curve are common with the two {\it NICER} bursts.}
	 \label{fig:burst_lc}}
\end{figure*}

 \begin{figure*}
\centering{
\includegraphics[width=5.45cm,angle=270]{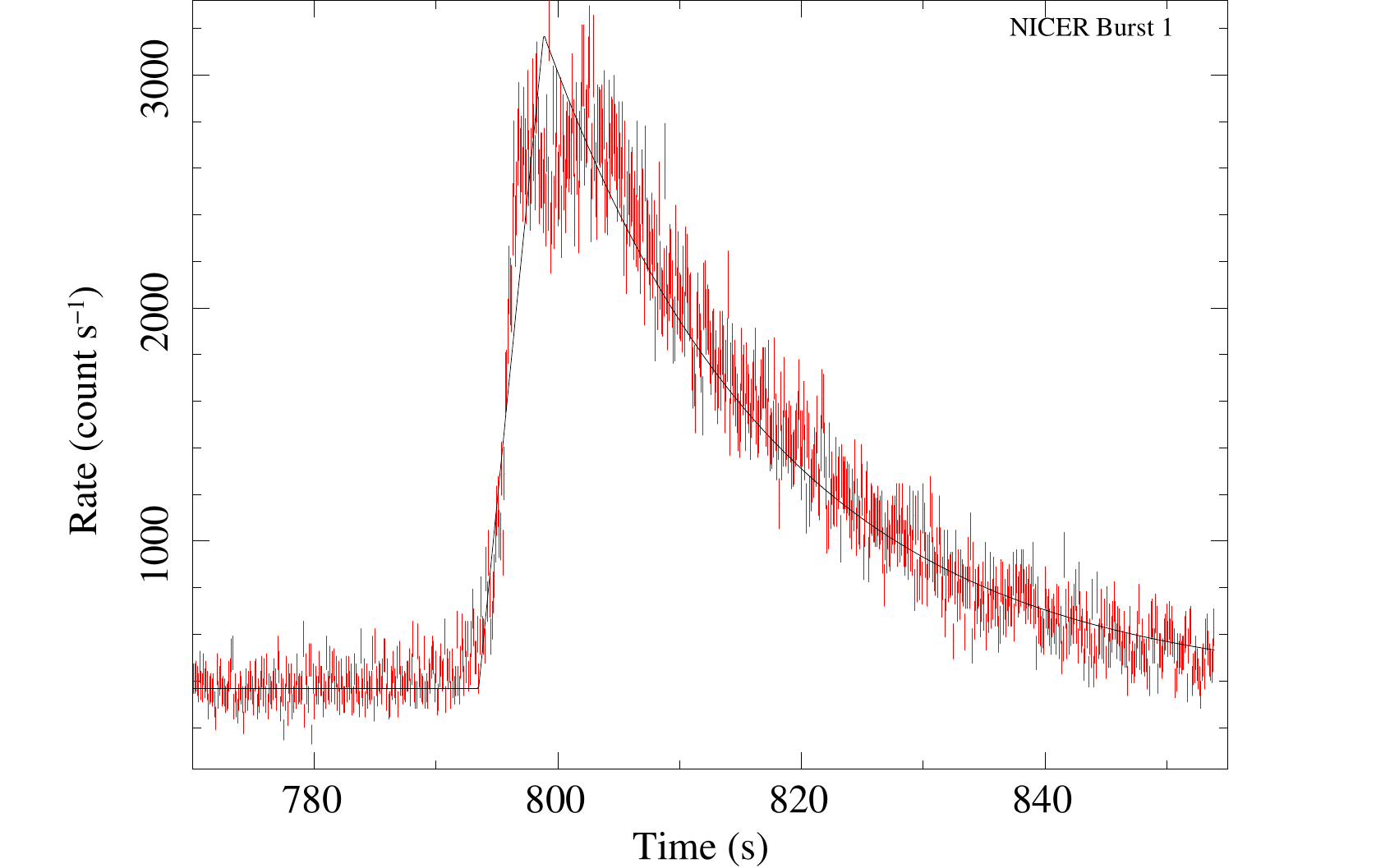}
\includegraphics[width=5.45cm,angle=270]{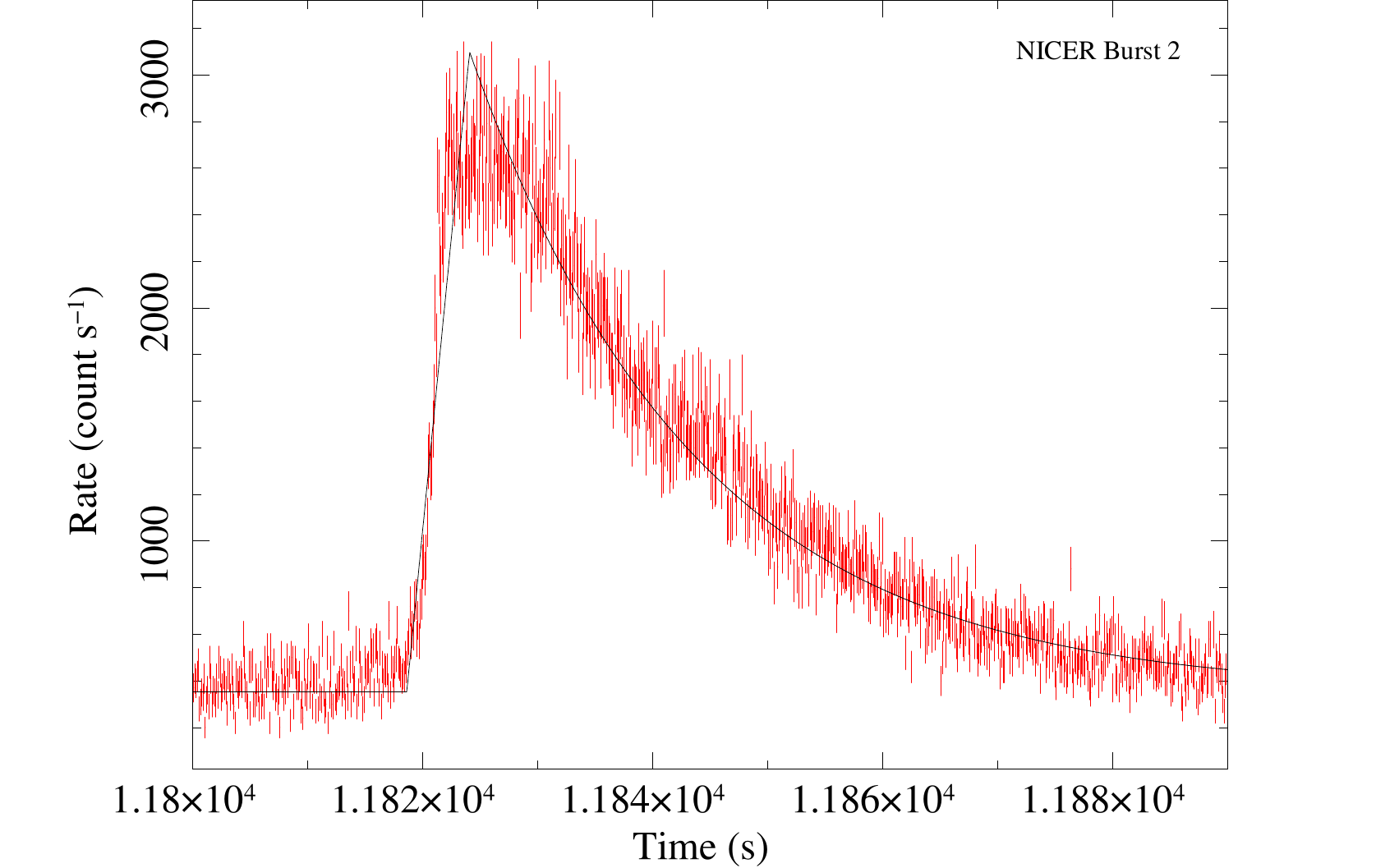}
\includegraphics[width=5.6 cm, angle=270]{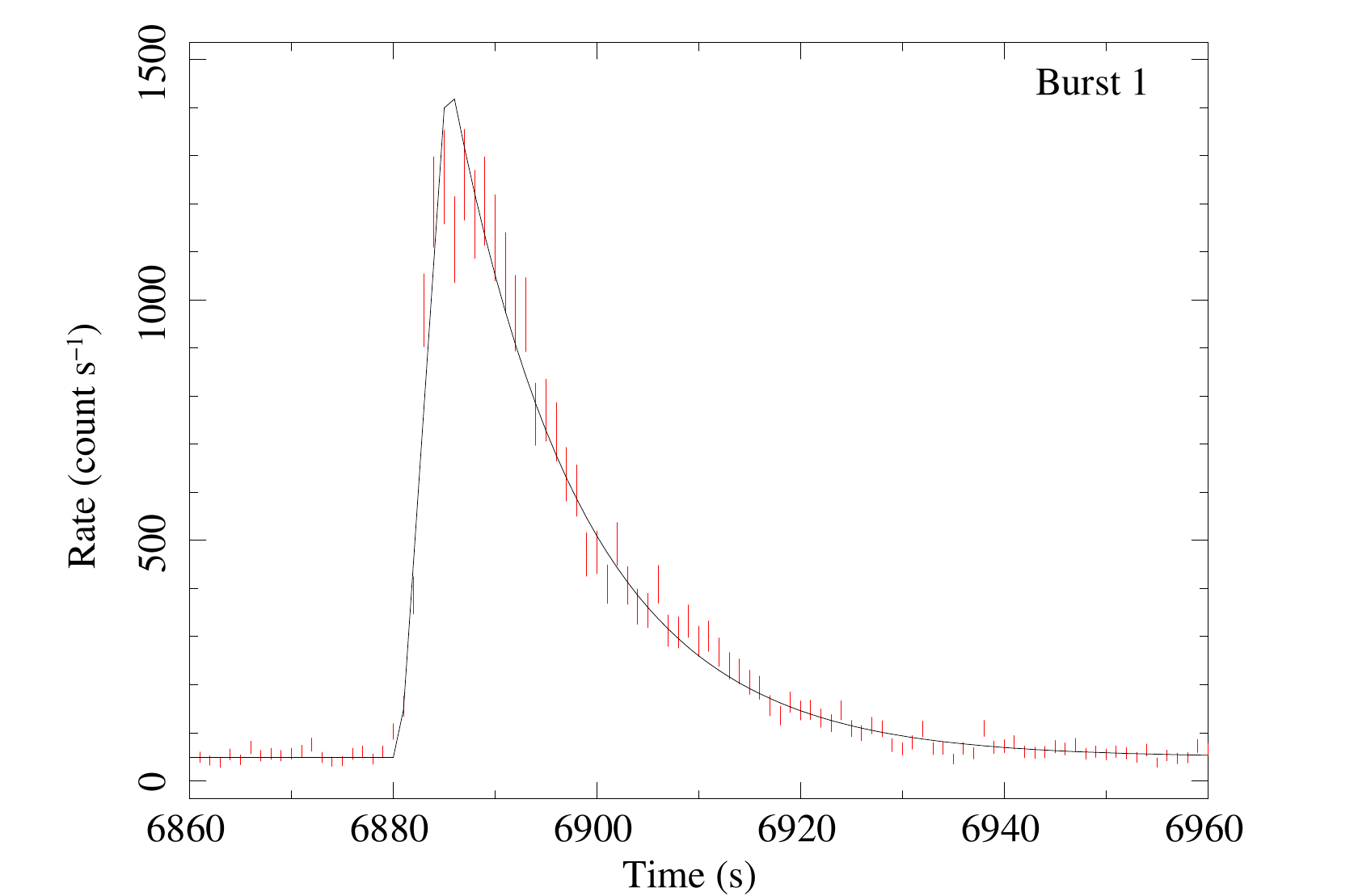}
\includegraphics[width=5.6 cm, angle=270]{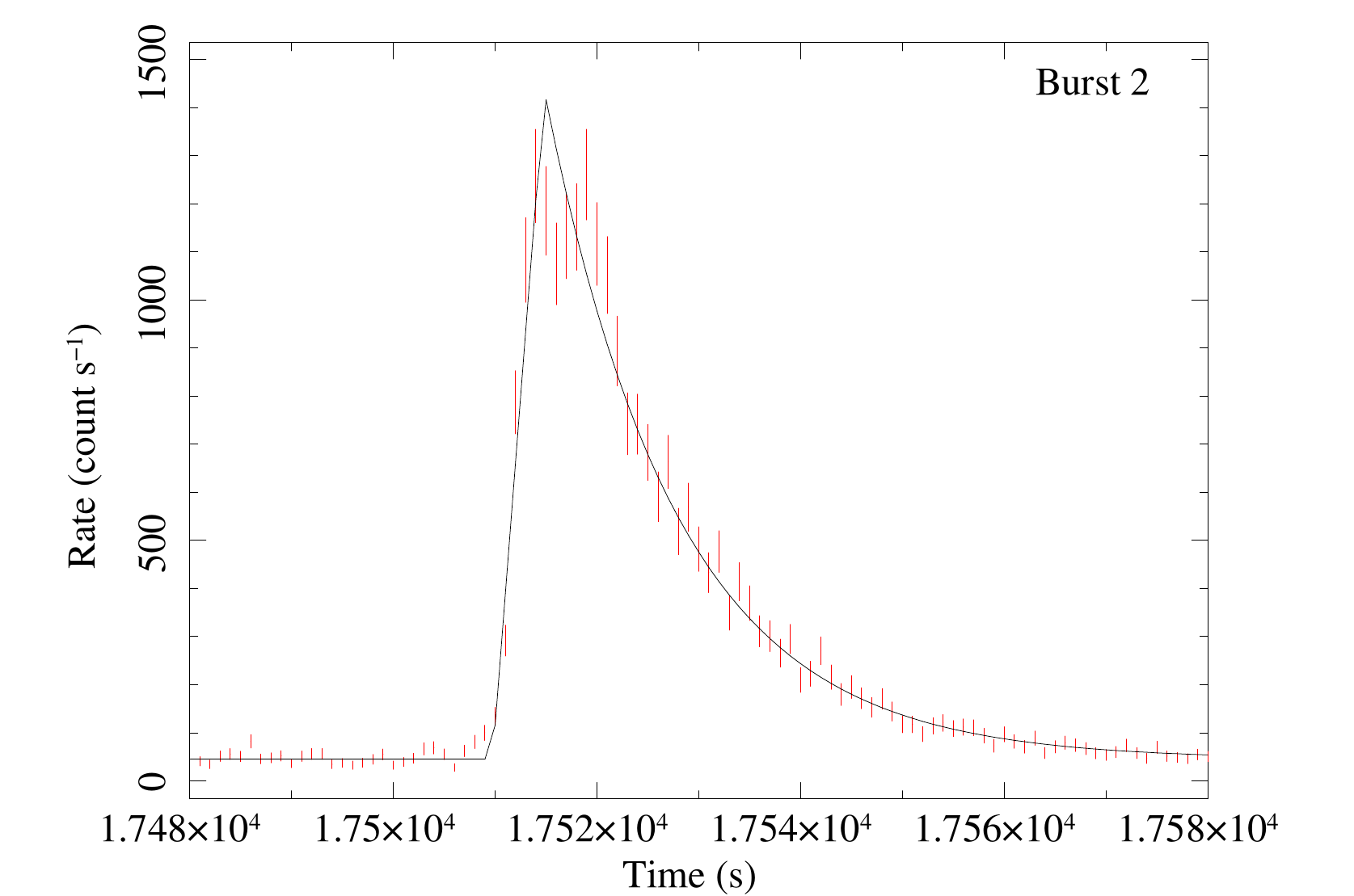}
\includegraphics[width=5.6 cm, angle=270]{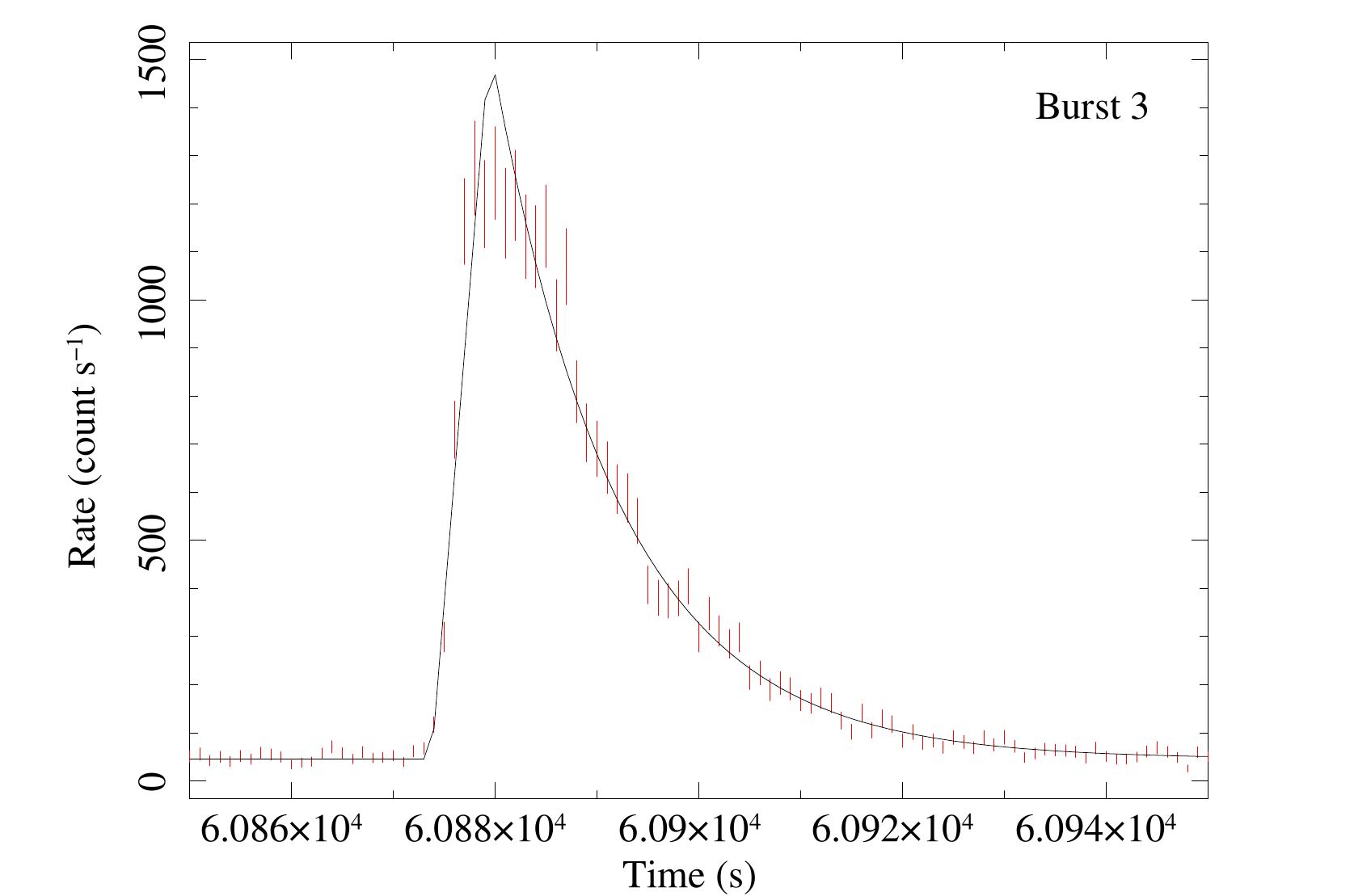}
\includegraphics[width=5.6 cm, angle=270]{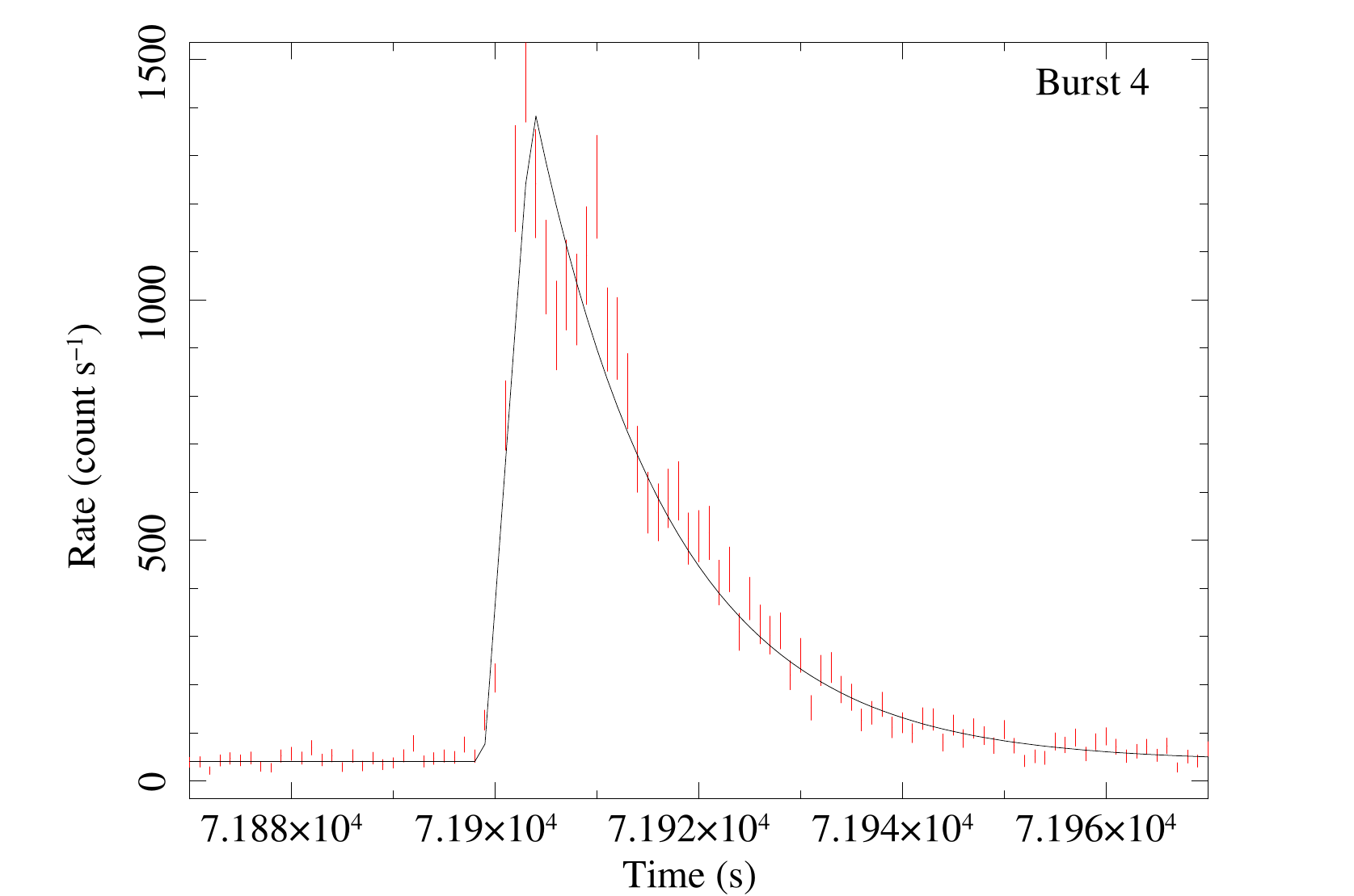}
\caption{Modelling of four X-ray bursts detected from MAXI J1616--195 using {\it NuSTAR} and {\it NICER}. The burst profiles are modelled with the {\tt burs} model (a linear rise and an exponential decay) and the best-fitting parameters are summarised in Table \ref{tab:log_table_burst_model}. Two thermonuclear bursts are detected from MAXI J1816--195 with {\it NICER} (0.4--10 keV) using 1 s binned light curve (Obs. ID--5533011601), shown in the top left and top right-side images of the figure. The rest of the four images represent thermonuclear bursts, which are detected from {\it NuSTAR} observation. The  duration of each burst is nearly 30 s. 
}

 \label{fig:burst_modelling}}
\end{figure*}

\begin{figure}
\centering{
\includegraphics[width=6cm, angle=270]{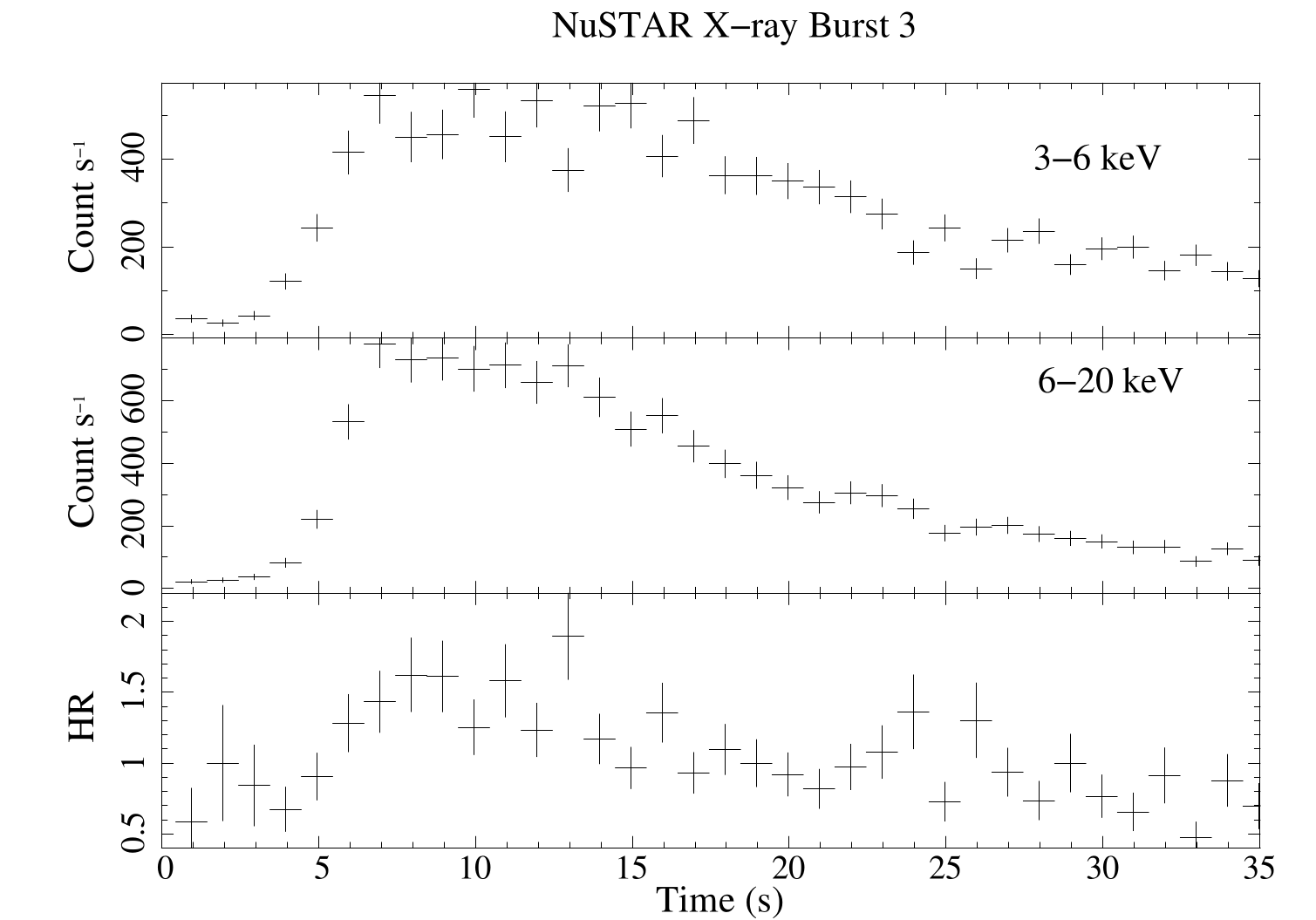}
\caption{{Evolution of hardness ratio of MAXI J1816--195 using {\it NuSTAR} during a thermonuclear burst.  {\it NuSTAR} detected a total of four bursts during the observation and this HR is shown for the third burst. The HR for the rest of the three bursts is shown in Fig. \ref{fig:Hardness_evolution}. The first and second panels show the 1 s binned light curve in the energy band of 3--6 keV and 6--20 keV, respectively. The bottom panel shows the variation of the hardness ratio (6--20 keV)/(3--6 keV). The hardness ratio evolved with the burst and reached a maximum value of $\sim$2 during the peak of the burst.}}
	 \label{fig:HR}}
\end{figure}

\section{Data analysis and methodology}
\label{obs}
We analyzed data taken by {\it NuSTAR} and {\it NICER} covering the entire duration of the outburst, starting from May 2022 and lasting for nearly two months. We reduced data from the two satellites using {\tt HEASOFT} version 6.28. We also used final data products (light curves) provided by {\it MAXI} and the Burst Alert Telescope onboard Neil Gehrels Swift Observatory\footnote{\url{https://swift.gsfc.nasa.gov/results/transients/}} \citep{Kr13} missions.

\subsection{{\it NuSTAR} observation}
The Nuclear Spectroscopic Telescope Array ({\it NuSTAR}) consists of two identical X-ray telescope modules that are co-aligned and operate in a broad energy range of 3--79 keV. Each detector of each telescope (typically known as focal plane modules A (FPMA) and B (FPMB)) provides a spectral energy resolution of 400 eV (FWHM) at 10 keV \citep{Ha13}.
{\it NuSTAR} performed an observation of MAXI J1816--198 on 2022 June 23--24 with an exposure of $\sim$35 ks. 
 The data were reduced using the {\tt NuSTARDAS pipeline} provided under {\tt HEASOFT} with {\tt CALDB} version of 20221130. The data was screened and calibrated using {\tt NUPIPELINE}. The source light curves and spectra were extracted using a circular region of radius 60 arcsec centred at the source position using {\tt NUPRODUCTS} scripts provided by the {\tt NuSTARDAS} pipeline. The background light curves and spectra were extracted from circular regions 100 arcsec away from the source. The background correction of the light curve was made using {\tt lcmath}.

\subsection{{\it NICER} observation}
A soft X-ray non-imaging spectroscopy and timing instrument, the Neutron Star Interior Composition Explorer ({\it NICER}) is located on the International Space Station. The main component of {\it NICER} is the X-ray Timing Instrument (XTI), which operates in the soft X-ray range of 0.2--12 keV \citep{Ge16}. A series of follow-up observations of MAXI J1816--195 was taken with {\it NICER} beginning in 2022 May. 
 We have used a single {\it NICER} observation (Obs. ID--5533011601) with exposure of $\sim$2.33 ks which was simultaneous with {\it NuSTAR} observation.
 {\tt NICERDAS} in {\tt HEASOFT} was used to process  the raw data. By applying the standard calibration and filtering tool  {\tt nicerl2} to the raw data, we produced clean event files. We used {\tt XSELECT} to extract light curves and spectra for MAXI J1816--195 from the clean event files. 
 We chose good time intervals for the timing analysis based on the following criteria: the ISS was not in the South Atlantic Anomaly (SAA) region, the source direction was at least 30$^{\circ}$ away from the bright Earth, and the source elevation was greater than 20$^{\circ}$ above the Earth's limb. We used the task  {\tt barycorr} to apply barycentric corrections for the time analysis. 
 Light curves with a time resolution of 1 s containing events in the 0.4--10 keV are generated for timing analysis. We have analyzed the {\it NICER} data with the {\tt CALDB} version of 20210707. The {\tt nibackgen3C50}\footnote{\url{https://heasarc.gsfc.nasa.gov/docs/nicer/tools/nicer_bkg_est_tools.html}} tool was used to simulate the background for each observational epoch \citep{Re22}.

\begin{figure*}
\centering{
\includegraphics[width=8.5cm]{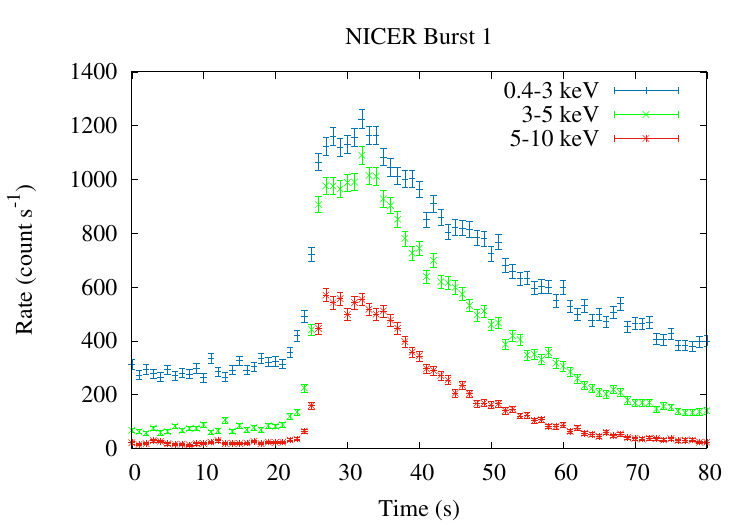}
\includegraphics[width=8.5cm]{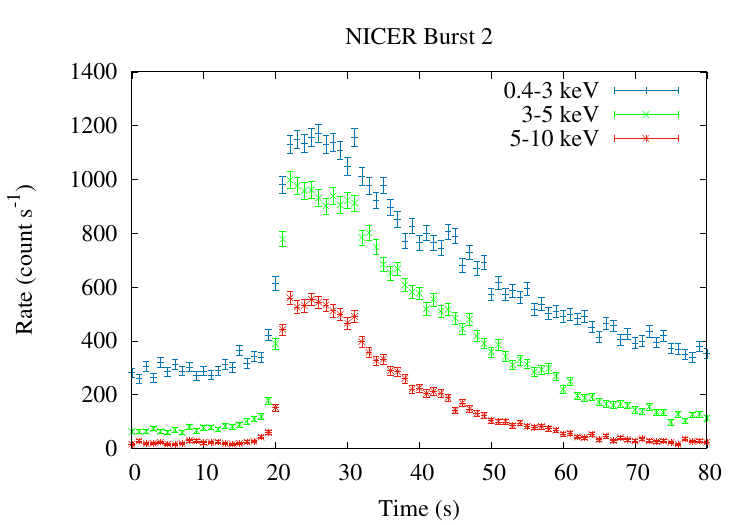}
\includegraphics[width=8.5 cm]{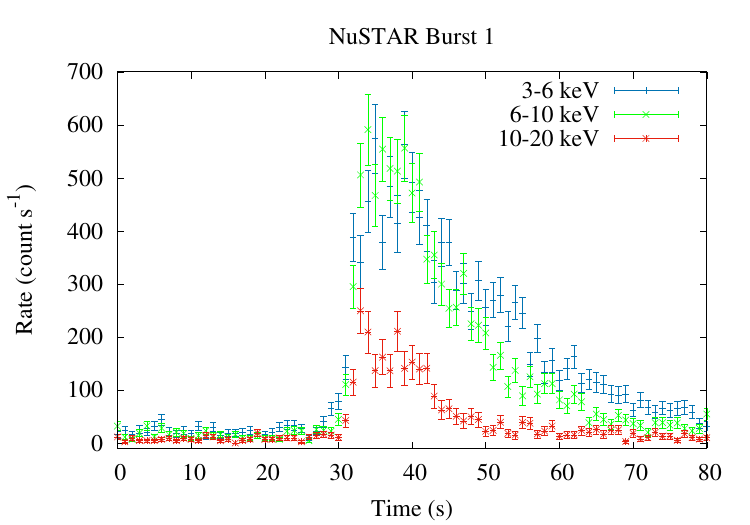}
\includegraphics[width=8.5 cm]{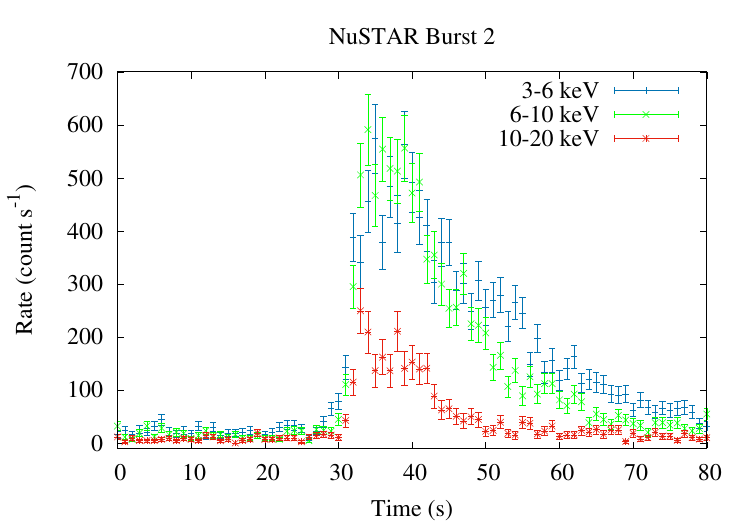}
\includegraphics[width=8.5 cm]{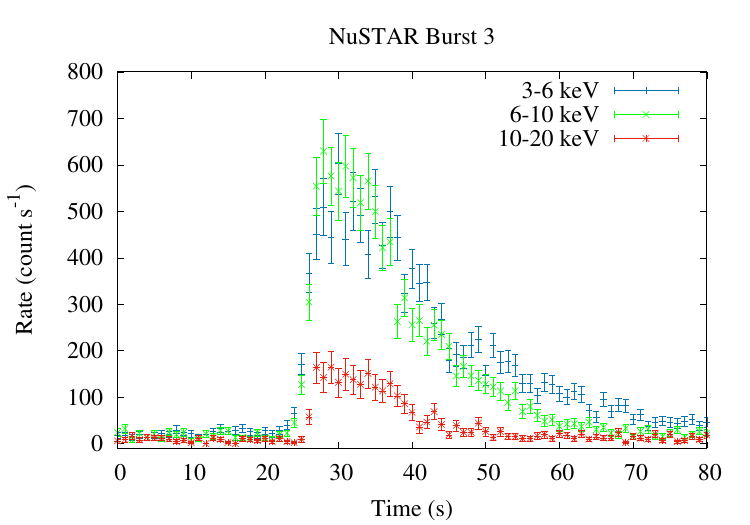}
\includegraphics[width=8.5 cm]{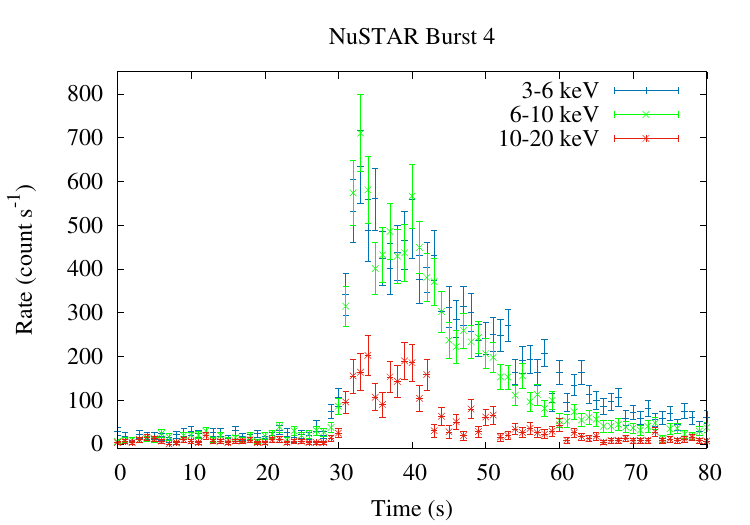}
\caption{Energy-resolved burst profiles are shown using {\it NICER} and {\it NuSTAR}/FPMA simultaneous observation data. The top left and right-hand side panels show energy-resolved burst profiles for MAXI J1816--195 using {\it NICER} with a 1-s binned light curve (Obs. ID--5533011601) and the rest of the bursts are observed using {\it NuSTAR}. Four thermonuclear bursts are detected from MAXI J1816--195 using {\it NuSTAR} 1 s binned light curves. The burst profiles show a comparatively lower peak count rate at higher energy bands. Each burst is continued for nearly 30 s. The bursts are detected significantly up to an energy of 20 keV using {\it NuSTAR}.}
	 \label{fig:burst_profile}}
\end{figure*}

\section{Results}
\label{res}
The millisecond pulsar MAXI J1816--195 was discovered by {\it MAXI} during an outburst in 2022 May. Several follow-up observations have been conducted to study the source's multi-wavelength behaviour \citep{Bu22,Ke22, Ch22, Ken22, de22, Br22, Be22}. For example, we used the  \textit{Swift}/BAT telescope's high energy X-ray coverage (15-50\,keV) and daily monitoring of the entire sky, to ascertain the spectral state of the source and duration of the outburst. Fig. \ref{fig:BAT} shows the outburst from MAXI J1816--195 during 2022 May--June as seen with {\it MAXI} and {\it BAT}. The outburst from MAXI J1816--195 continued for nearly 30 days. The maximum flux reached by the source was $\sim$0.15 crab as observed with {\it Swift}/BAT (15--50 keV) on 2022 June 13 (MJD 59743). Simultaneous observation with {\it NuSTAR} and {\it NICER} was conducted on 2022 June 24 and is the focus of this paper. The observation time is indicated using two arrows in Fig. \ref{fig:BAT}. It is clear from the figure, that the simultaneous {\it NICER/NuSTAR} observation was performed in the decay phase of the outburst.
The source was detected using {\it NuSTAR} at the location of RA = $18^h 16^m 52^s.40$, Dec = $-19^{\degr}37^{'} 58^{''}$.35. and the source is 1.45 arcsec away from the {\it Swift}/XRT localization  \citep{Ke22}, which is within the 2.2 arcsec (90 per cent confidence) error radius of the XRT position of the source. 

\begin{table}
\centering
\caption{Details of X-ray bursts detected from MAXI J1816--195 using {\it NuSTAR} and {\it NICER} observations. 
}
\label{tab:log_table_burst}
\begin{tabular}{cccc} 
	\hline
	Instrument & Start time       & Number of bursts  & Obs. ID \\
		   &   (MJD)          &   &                    \\
\hline

{\it NuSTAR} &  59753.45 &  4 & 90801315001  \\
&    &    &   \\
\hline
{\it NICER}		& 59754.14 &  2 & 5533011601\\

\hline
	\end{tabular}
\end{table}

\subsection{X-ray bursts and burst timing analysis}
Multiple thermonuclear bursts are found from MAXI J1816--195 during the outburst of 2022 using {\it NuSTAR} and {\it NICER}.
The X-ray bursts were detected by the {\it NuSTAR} observation on 2022 June 23. The left-hand side panel of Fig. \ref{fig:burst_lc} shows the light curve (3--79 keV) using {\it NuSTAR}/FPMA, which shows four bursts during the observation. The average peak count rate of the bursts was estimated as $\sim$1300 counts s$^{-1}$ from NuSTAR/FPMA. The right-side panel of Fig. \ref{fig:burst_lc} shows the light curve of the source using {\it NICER}, which indicates two thermonuclear bursts with a peak count rate of $\sim$ 3000 counts s$^{-1}$ during simultaneous observation with {\it NuSTAR}. The summary of detected bursts from MAXI J1816--195 is given in Table \ref{tab:log_table_burst}.   
The peak count of burst 1, burst 2, and burst 3 was $\sim$ 1260 counts s$^{-1}$and the duration of the bursts was $\sim$ 30 s. For burst 4, the peak count was $\sim$ 1500 counts s$^{-1}$ with a burst duration of $\sim$ 30 s. The right-hand side image of Fig. \ref{fig:burst_lc} shows the light curve using {\it NICER}, which is simultaneous with {\it NuSTAR}. The simultaneous observations of the source using {\it NICER} (Obs. ID–533011601) and {\it NuSTAR} (Obs. ID-90801315001) implies that the burst 3 and burst 4 of {\it NuSTAR} are simultaneous with {\it NICER} bursts (Fig. \ref{fig:burst_lc}). For the {\it NuSTAR} observation, the recurrence time between bursts 1 and 2 was roughly 10.64 ks, and the recurrence time between bursts 3 and 4 was approximately 11.03 ks (the data gap between two consecutive bursts was $\sim$5.5 ks). The recurrence time for two successive bursts for the {\it NICER} observation was $\sim$11 ks (the data gap between two consecutive bursts was $\sim$10 ks).

The {\it NuSTAR} burst profiles are modelled using the {\tt QDP} model {\tt burs} shown in Fig. \ref{fig:burst_modelling}. The characteristics of Type-I bursts can be modelled using the fast linear rise and exponential decay profile. The burst profiles are fitted with a constant and the burst model (linear rise and exponential decay).
The {\it NuSTAR} X-ray burst fitting parameters imply a sharp rise time (4.6 s to 7.6 s) and a slow decay duration of 12.3 to 13.4 s. The X-ray bursts detected using {\it NuSTAR} indicate that the peak count rates are around 26 times higher than the persistent levels. The {\it NICER} burst profiles are also fitted with the same model {\tt burs} and the best-fitting parameters are summarised in Table \ref{tab:log_table_burst_model}. Fig. \ref{fig:burst_modelling} shows two bursts using {\it NICER} data in the 0.4--10 keV energy band during the simultaneous observation with {\it NuSTAR}. The duration of the burst was nearly 30 s, and the pattern was a sharp linear rise in flux and comparatively slow exponential decay. 
 The NICER X-ray burst fitting parameters for {\it NuSTAR} indicate a rapid rise time of 5.5–5.8 s and a gradual decline time of $\sim$19 s. The burst duration was around 35 s. The peak count rates are $\sim$ 7.5 times higher than the persistent levels.

\subsection{Evolution of hardness ratio}
We begin by studying the variation of the hardness ratio during the thermonuclear bursts using {\it NuSTAR}. The light curves in two energy bands, 3--6 keV and 6--20 keV, are produced during the burst with a time resolution of 1\,s. The bottom panel of Fig. \ref{fig:HR} shows the evolution of the hardness ratio (6--20 keV)/(3--6 keV) during the thermonuclear burst as observed with \textit{NuSTAR}.  The HR is shown for the {\it NuSTAR} burst-3. Initially, the hardness increases from 0.5 to a peak of about 1.5. The HR gradually decreases and returns to a persistent level, as the burst flux begins to decrease. The hardness ratio shows a similar pattern for all four X-ray bursts during the {\it NuSTAR} observation.

\begin{table*}
\centering
\caption{Fitting parameters of X-ray bursts using the {\tt QDP} model {\tt burs} from simultaneous {\it NuSTAR} observation (Obs. ID - 90801315001) and {\it NICER} observation (Obs. ID - 5533011601).}
\label{tab:log_table_burst_model}
\begin{tabular}{ccccccc} 
	\hline
	Burst No. & Burst start time  & Burst duration       & Peak to persistent count rate & Duration of rising &  Duration of decay & Peak count rate\\
		{\it NuSTAR}   &   (s)          &(s) &  & (s) & (s)  & (count s$^{-1})$                \\
\hline

  1  & 6880   & $\sim$30  & $\sim$26.2 & $\sim$4.6 & $12.8\pm 0.5$ & $1259\pm 91$ \\
  2  & 17510  & $\sim$30  & $\sim$26.1 & $\sim$7.6 & $12.5\pm 0.1$ &  $1256\pm 94$ \\
  3  & 60875  & $\sim$30   & $\sim$26.5 & $\sim$5.7 & $12.3\pm 0.4$ & $1273\pm 98$ \\
  4 & 71900  &  $\sim$25   & $\sim$31.3 & $\sim$4.7 & $13.4\pm 0.5$ & $1503\pm 121$ \\
\hline
{\it NICER} burst  1  &  795 & $\sim$35  & $\sim$7.7 & $\sim$5.8 & $19.1\pm 0.8$ & $3100\pm 60$ \\
{\it NICER} burst  2  & 11820 & $\sim$35  & $\sim$7.4 & $\sim$5.5 & $19.6\pm 0.8$ &  $2970\pm 40$ \\
\hline

	\end{tabular}
\end{table*}

\subsection{Energy-resolved burst profiles}
We have studied the energy dependence of the burst profiles of MAXI J1816--195 using simultaneous {\it NuSTAR} and {\it NICER} data during the outburst. 
The top left and top right-side images of Fig. \ref{fig:burst_profile} show the energy-dependent burst profiles using {\it NICER}. The burst profiles are generated for the energy ranges of 0.4--3 keV, 3--5 keV, and 5--10 keV for two different bursts during the {\it NICER} observation. Fig. \ref{fig:burst_profile} shows that the peak count rate decreases with an increase in energy, and the peak count rate in the higher energy band 5--10 keV is $\sim$2.5 times less compared to 0.4--3 keV energy band. The {\it NICER} burst profiles significantly evolved with energy, and during the decay phase of the burst, a comparatively long tail is seen in the burst profile at low energy.

The energy dependence of the burst profiles is also observed from the {\it NuSTAR} observation. Fig. \ref{fig:burst_profile} shows burst profiles at different energies (3--6 keV, 6--10 keV, 10--20 keV) using {\it NuSTAR}/FPMA data. The X-ray bursts are detected in different energy bands of {\it NuSTAR}. It is visible that the peak fluxes of bursts evolve with energy, and with increasing energy, the amplitude of the burst peak count rate decreases. The energy-resolved bursts are significantly detected up to 20 keV using {\it NuSTAR}. The peak count rate of the burst profile is reduced by a factor of $\sim$4 in the higher energy band (10--20 keV) compared to the lower energy band (3--6 keV). A relatively large tail is found in the low energy range (3--6 keV) compared to the high energy range (10--20 keV) during the burst decay phase. The variation of the hardness ratio is also investigated to look for the evolution of the source state during the burst. Fig. \ref{fig:HR} represents the evolution of the hardness ratio for burst-3 using the {\it NuSTAR} observation. The variation of the hardness ratio shows a similar trend for all four bursts during {\it NuSTAR} observation. During the rising phase of the burst, hard photons (6--20 keV) dominate the soft band photons (3--6 keV) as HR increases till the peak of the burst. In the decay phase of the burst, the HR decreases and the soft photons dominate over the hard photons.

\begin{figure*}
    \includegraphics[width=1.0\columnwidth]{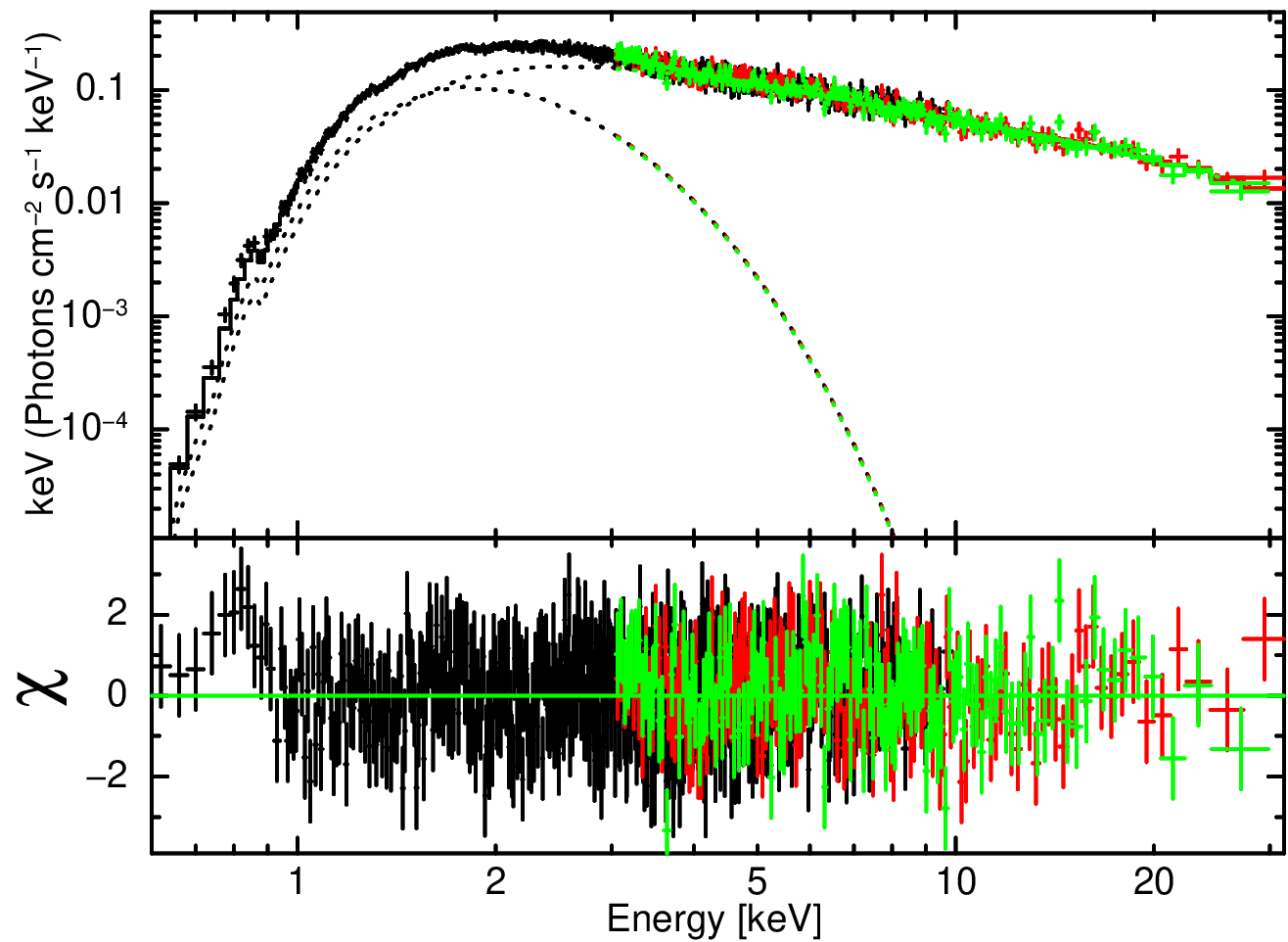}
    \includegraphics[width=1.0\columnwidth]{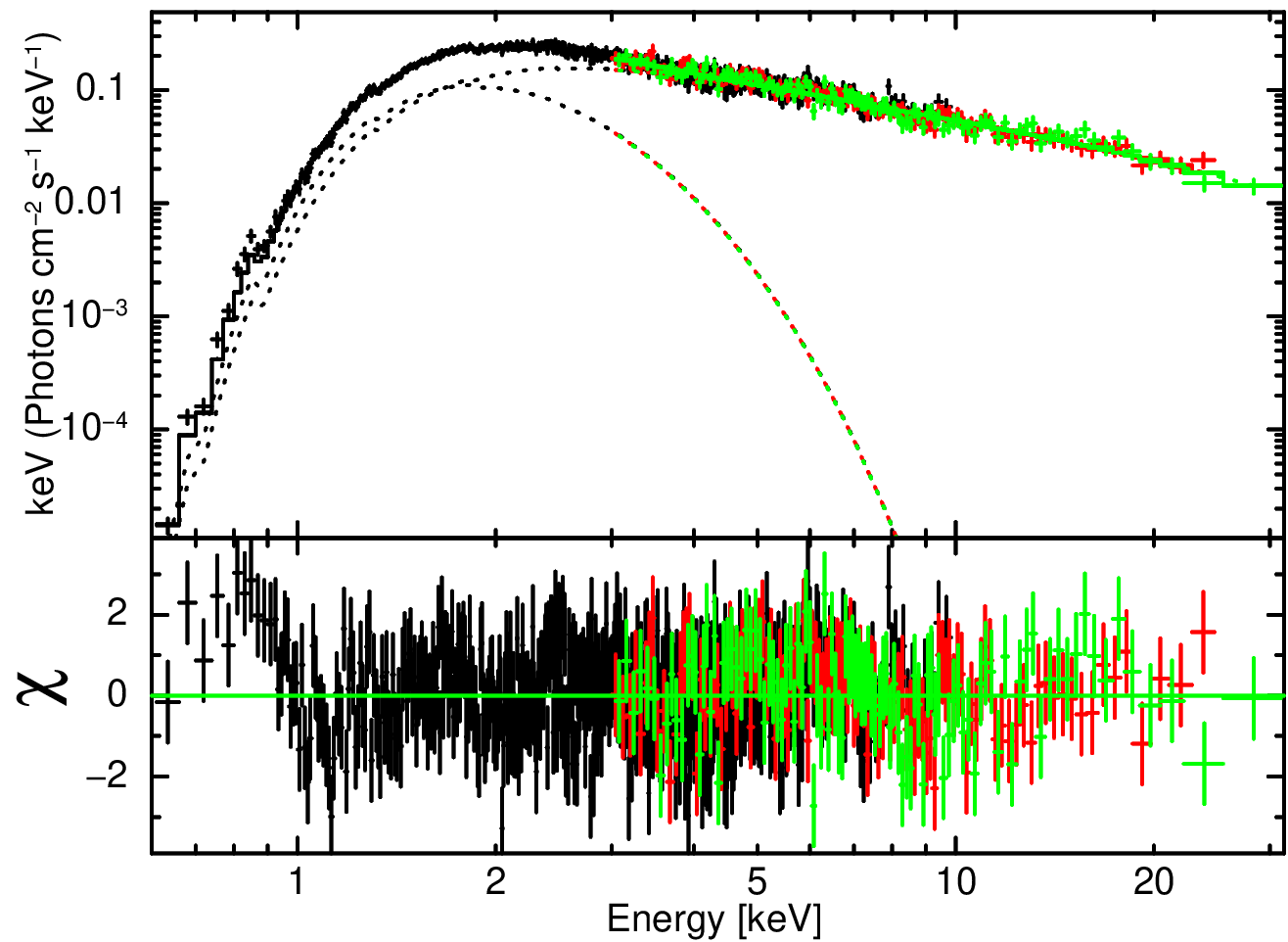}
    \caption{The plots highlight the fitted spectra using simultaneous {\it NICER/NuSTAR} data of the thermonuclear burst-3 and burst-4 for the initial persistent time bin P1. The best-fit model is {\tt TBabs$\times$(diskbb + nthComp)}. The fitting model and strategy are discussed in Section \ref{time_resolved_spectroscopy}. The best-fitting parameters are presented in Table \ref{tab:tab3}.}
\label{fig:spec_burst3}
    \label{fig:appendix_fig2}
\end{figure*}

\subsection{Time-resolved burst spectroscopy}
\label{time_resolved_spectroscopy}
To understand the dynamic evolution of observed thermonuclear bursts, we performed time-resolved  spectroscopy and studied the variation of different spectral parameters during the burst. The pre and post-burst spectra for MAXI J1816--195 are obtained from the joint analysis of the {\it NICER} and {\it NuSTAR} data and modelled in {\tt XSPEC} \citep{Ar96}. The joint pre and post-burst spectra (0.5--79\,keV) are obtained for a time bin of $\sim$300 s. We used only {\it NICER} data (0.5--10.0\,keV) to extract time-resolved burst spectra. A total of 25 spectra (P1 to P25) are generated for each burst, P1 and P25 are the pre and post-burst spectra, and the rest of them (P2 to P24) are during the burst. Two different time bins are taken to extract the time-resolved spectra. We selected a time bin of  2\,s to extract the spectra from P2 to P18 and a time bin of  4\,s from P18 onward. To optimize the signal-to-noise ratio in the burst's decaying portion, we extended the time bin from 2\,s to 4\,s.

The pre and post-burst spectra are described by the thermal Comptonization {\tt XSPEC} model {\tt nthComp} \citep{Zd96, Zy99} along with a disc ({\tt diskbb}) component. The interstellar medium absorption is accounted for by the {\tt TBABS} \citep{Wi20} model in {\tt XSPEC}. The best-fit broadband pre-burst spectra are shown in Fig. \ref{fig:spec_burst3} using the simultaneous observation of {\it NuSTAR} and {\it NICER} during burst 3 and burst 4 respectively. The results of the pre and post-burst spectral fitting are summarised in Table \ref{tab:tab3}. The average hydrogen column density ($N_H$) from the pre and post burst spectra was $\sim 2.38 \times 10^{22}$ cm$^{-2}$. The photon index shows a consistent value of $\sim2$ for the pre-burst and post-burst intervals. The joint spectral modelling provides an electron temperature of $\sim11$ keV for the P1 phase of each burst. The lower value of the electron temperature during the pre-burst time implies the source is in the soft state. 

  We divide both the bursts into 23-time bins, namely; P2 to P24. The P2 and P3 time intervals correspond to the sharp rise, whereas time bin P4 signifies the peak of the  thermonuclear burst (TNB). We further broke the exponential decay part of the TNB into 19 parts, which are represented by the time bins P5 to P24. We have selected a time bin of 2 s for the first 17 points (P2 to P18) and the time bin increased to 4 s for the rest of the burst (P19 to P24). For each interval, the spectra are extracted through {\tt XSELECT} using the good time interval files (GTIs) for each burst for NICER data. 
 
\begin{figure*}
    \includegraphics[width=2.1\columnwidth]{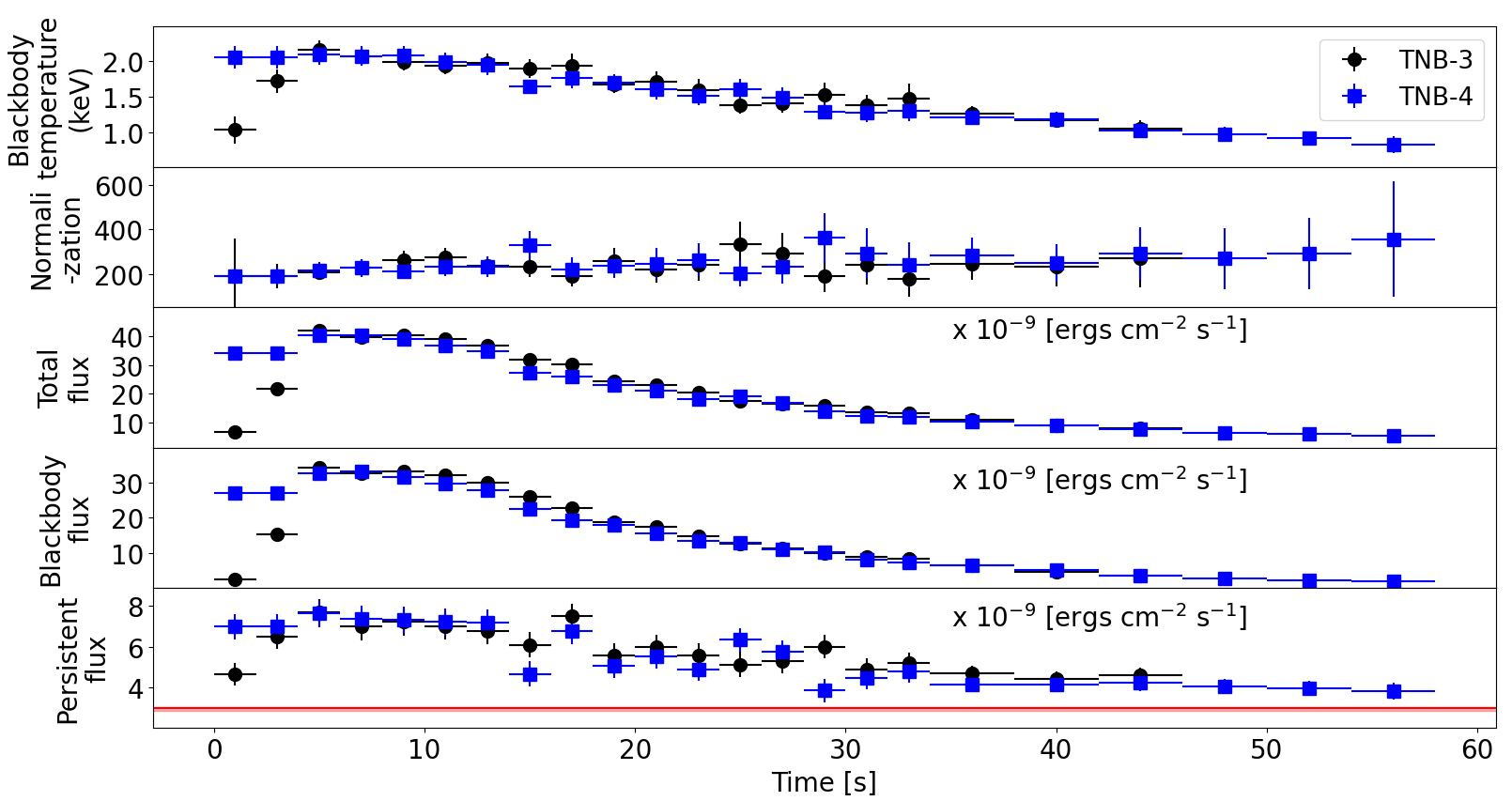} \vspace{-0.6cm}

\caption{The evolution of the various spectral parameters  during the evolution of the TNB: blackbody temperature (kT; Top panel), blackbody normalization (second panel), total flux (third panel), blackbody flux (fourth panel), and persistent flux (bottom panel). The X-axis presents the time from the rise of the TNB. The red horizontal line in the bottom panel highlights the average persistent flux (non-thermal) during the pre-burst time.}
    \label{fig:fig_time_Resolved}
\end{figure*}

The {\it NICER} time-resolved spectra were modelled using the combination of absorbed blackbody along with a non-thermal component to account for the persistent emission. 
 The burst spectra are fitted with a combination of galactic absorption, a blackbody ({\tt bbodyrad}) model, and the persistent spectrum during the burst is accounted for using the $f_a$ method \citep{Worpel2013}. The best-fitting results of time-resolved  spectra are summarised in Table \ref{tab:tab4}. During the decay phase (P22 to P24) of the TNB 3, the signal-to-noise ratio was low and fitting statistics were not good, we have not added this in Table \ref{tab:tab4}. The hydrogen column density is fixed to the average value of pre-burst and post-burst level $\sim 2.38\times 10^{22}$ cm$^{-2}$ during the fitting of the {\it NICER} time-resolved spectra. The dimensionless $f_a$ parameter evolves during the burst as the persistent emission shows variation during this time \citep{Worpel2013}. We look for the variation of the persistent emission during the burst through the $f_a$ method. We have applied the $f_a$ method to account for the non-thermal emission for the time-resolved spectra (P2--P24). The $f_a$ value does not show significant variation during the burst and $f_a$ reached a maximum of $1.74\pm0.21$ for burst 3 at the peak of the burst (P4 phase). Compared to traditional methods, the flux value in the $f_a$ method is lower around the peak of the burst. The variation of different spectral parameters during the burst is shown in Fig. \ref{fig:fig_time_Resolved}. The top panel of Fig. \ref{fig:fig_time_Resolved} shows the variation of the blackbody temperature; the second panel shows the variation of the blackbody normalization; the third and fourth panels show the evolution of the total flux and blackbody flux, respectively. The bottom panel of Fig. \ref{fig:fig_time_Resolved} shows the evolution of the persistent flux of the source. The blackbody temperature varies between 1.03\,keV, and 2.16 keV during burst 3 and for burst 4, kT varies between 0.82 keV to 2.09 keV. The blackbody temperature attains its maximum value of 2.16\,keV during the peak (P4) of the burst and the normalization of the blackbody was $\sim207$ for burst 3. From the normalization, the blackbody radius ($R_\mathrm{bb}$) can be estimated via,  $\mathrm{Norm}_\mathrm{bb} = \frac{R_\mathrm{bb}^{2}}{D_{10 }^{2}}$, where $D_{10}$ is the source distance in units of 10\,kpc.  
A peak flux of $\sim 4.2 \times 10^{-8}$ erg cm$^{-2}$ s$^{-1}$ is reached during the P4 phase of burst 3. 
The apparent emitting area of the blackbody is estimated from the average normalization of the blackbody component for burst 3 and burst 4 during the P4 phase. The apparent emitting area of the blackbody corresponding to the normalization $\sim$212 is estimated to be $\sim$9.2 km (assuming the source distance is $\sim$6.3 kpc \citep{Che22}) and $\sim$12.5 km (assuming the source distance is $\sim$8.6 kpc \citep{Bu22}) respectively.

\begin{table*}
\centering
 \caption{Best-fitting parameter values [{\tt XSPEC} model $\texttt{TBabs}\times(\texttt{diskbb}\,\times\,\texttt{nthComp})$] of the $\sim300$\,s pre-burst and post-burst {\it NICER} and {\it NuSTAR} joint spectra of burst\,3 and burst\,4 from MAXI J1816--195.}
\begin{center}
\scalebox{1.2}{%
\begin{tabular}{ |l|l|c|c| }
\hline
\hline
Components & Parameters & Pre-burst & Post-burst \\
 &  & P1 & P25 \\
  &  & (373 s) & (242 s)  \\
 
\hline
\hline
   & Burst 3 &  &  \\
\hline
\hline
TBABS & N$_\mathrm{H}$ ($\times 10^{22} cm^{-2}$) & 2.39$\pm0.04$ & 2.36$\pm0.56$ \\
\hline
diskbb & T$_\mathrm{in}$(keV) & 0.5$\pm0.02$ & 0.7$\pm0.25$ \\
  & Norm. & 1428$^{+296}_{-231}$ & 237$^{+1857}_{-145}$ \\
\hline
nthComp & $\Gamma$ & 2.0$\pm0.04$ & 2.1$^{+0.03}_{-0.25}$ \\
  & kT$_{e}$(keV) & 11.2$^{+8.9}_{-2.6}$ & 1000$^{+}_{-}$ \\
  & Norm. ($\times$ 10$^{-1}$) & 4.0$\pm0.3$ & 5.7$^{+1.8}_{-3.0}$ \\
\hline
 & F$_\mathrm{Total}$ ($\times$ 10$^{-9}$ ergs cm$^{-2}$ s$^{-1}$) & 4.3$\pm0.04$ & 6.1$\pm0.60$ \\
 & F$_\mathrm{diskbb}$ ($\times$ 10$^{-9}$ ergs cm$^{-2}$ s$^{-1}$) & 1.4$\pm0.08$ & 0.6$\pm0.30$ \\
 & F$_\mathrm{nthComp}$ ($\times$ 10$^{-9}$ ergs cm$^{-2}$ s$^{-1}$) & 3.0$\pm0.06$ & 5.2$\pm0.70$ \\
\hline
 & $\chi^{2}$/dof & 803.13/799 & 210.30/209 \\
\hline
   \\
\hline
\hline
    &  Burst 4 & P1 & P25 \\
    &  & (350 s) & (322 s)  \\
\hline
\hline
TBABS & N$_\mathrm{H}$ ($\times 10^{22} cm^{-2}$) & 2.42$\pm0.04$ & 2.37$\pm0.07$ \\
\hline
diskbb & T$_\mathrm{in}$(keV) & 0.5$\pm0.02$ & 0.5$\pm0.03$ \\
  & Norm. & 1440$^{+306}_{-237}$ & 990$^{+330}_{-225}$ \\
\hline
nthComp & $\Gamma$ & 2.0$\pm0.05$ & 2.1$\pm0.05$ \\
  & kT$_{e}$(keV) & 11.7$^{+21.48}_{-3.29}$ & 32.3$^{+}_{-21.05}$ \\
  & Norm. ($\times$ 10$^{-1}$) & 3.9$\pm0.4$ & 4.3$\pm0.54$ \\
\hline
 & F$_\mathrm{Total}$ ($\times$ 10$^{-9}$ ergs cm$^{-2}$ s$^{-1}$) & 4.3$\pm0.05$ & 4.6$\pm0.08$ \\
 & F$_\mathrm{diskbb}$ ($\times$ 10$^{-9}$ ergs cm$^{-2}$ s$^{-1}$) & 1.4$\pm0.08$ & 1.2$\pm0.11$ \\
 & F$_\mathrm{nthComp}$ ($\times$ 10$^{-9}$ ergs cm$^{-2}$ s$^{-1}$) & 2.9$\pm0.07$ & 3.2$\pm0.12$ \\
\hline
 & $\chi^{2}$/dof & 745.65/719 & 622.81/585 \\
\hline
\hline
\label{tab:tab3}
\end{tabular}}\\
{\bf *}: All the errors are 90\% significant and calculated using the MCMC approach in {\tt XSPEC}\\
{\bf **}: All the flux values are unabsorbed and calculated for the energy band 0.5--79 keV
\end{center}
\end{table*}

\section{Discussion}
\label{dis}
We present the timing and spectral study of the newly discovered millisecond pulsar MAXI J1816--195 using {\it NuSTAR} and {\it NICER} data. The {\it NuSTAR} light curve shows four X-ray bursts and the duration of each burst was around 30 s.  At higher energies, the burst decays faster, which suggests that the temperature is falling as the burst evolves. Earlier, a similar type of trend was observed for several other sources, for example, 4U 1636--536 and GX 3+1 \citep{Be19, Na22}. The burst profiles for MAXI J1816--195 depend strongly on energy, and during the decay phase of the burst, a relatively long tail is observed in the burst profile at lower energies. The energy dependence of the burst profiles implies a softening of X-ray burst spectra in the decay phase of the burst due to the cooling of the photosphere of the neutron star \citep{Le93}. This is evident from the hardness ratio's significant evolution. During the rising phase, the hardness ratio increases until it reaches its maximum near the peak of the outburst and has started to decrease in the decay phase of the burst and returns to the persistent level.

The burst profile shows a sharp rise and comparatively slow exponential decay. The {\it NuSTAR} burst profiles indicate that the peak flux in 3--79 keV increased by a factor of $\sim$26 compared to the persistent level during the bursts. The energy resolved burst profiles for {\it NICER} data are estimated for the energies of 0.4--3 keV, 3--5 keV, and 5--10 keV (Fig. \ref{fig:burst_profile}), which show that the burst peak count rate reached nearly 1200 counts s$^{-1}$ in the 0.4--3 keV energy band and above 3 keV the burst peak count rates decay as expected for the Type-I X-ray burst. Earlier, Cyg X--2 showed that the peak count rate was maximum in the energy band of 3--6 keV, and above this, the count rates started to decrease, which suggests a Type-I X-ray burst \citep{De21}. The burst decay time gradually decreases with the increase in energy, which implies a decrease in temperature due to the cooling of burning ashes with the decay of the burst \citep{De16}.  
The {\it NuSTAR} energy-resolved burst profiles are generated for energy ranges of 3--6 keV, 6--10 keV, and 10--20 keV (Fig. \ref{fig:burst_profile}), which show that the count rate reached 600 counts s$^{-1}$ in the energy band of 3--6 keV, and beyond 6 keV, the peak count rate has started to decrease. In the higher energy band (10--20 keV), the peak count rate decreased to $\sim$ 200 counts s$^{-1}$. The burst decays faster in a higher energy range, which implies that the temperature decrease as the burst evolves. 

Time-resolved spectroscopy during the X-ray burst is crucial as different spectral parameters show significant evolution during the burst. Time-resolved spectroscopy was performed earlier for several sources to understand the accretion mechanism and variation of several spectral parameters during the burst. The time-resolved spectra for MAXI J1816--195 are described using a blackbody and an absorption component along with a non-thermal component. The blackbody temperature reached a maximum of $\sim$2 keV near the peak of the burst.
 The apparent emitting area of the blackbody has reached a maximum of $\sim$ 9.2 km and $\sim$12.5 km at the peak of the burst assuming a source distance of 6.3 kpc and 8.6 kpc respectively. Earlier, the MAXI J1816--195 was monitored by {\it Insight}/HXMT, and a detailed study was conducted by \citet{Che22}. Using {\it Insight}/HXMT observation of the source during the time the source was brightest, the photospheric radius expansion was not evident and the peak flux was estimated to be $\sim$8.03 $\times 10^{-8}$ erg cm$^{-2}$ s$^{-1}$ and the temperature and apparent emitting area of the blackbody reached the highest value of $\sim$2 keV and $\sim$11 km respectively \citep{Che22}. An upper limit of source distance of 6.3\,kpc was given by \citet{Che22} assuming the Eddington luminosity of 3.8 $\times$ 10$^{38}$ erg s$^{-1}$ \citep{Ku03}. Similarly, the time-resolved spectral analysis from {\it NICER} data for this source revealed that the expansion of photosphere expansion was not evident as the typical blackbody temperature and normalization were $\sim$1.9 keV and $\sim$300 at 10 kpc respectively \citep{Bu22}.
The blackbody flux varies between $\sim(0.2-3.4) \times 10^{-8}$ erg cm$^{-2}$ s$^{-1}$ during the bursts. The $f_a$ parameter varies between 1--1.7 and 0.8--1.7 for burst 3 and burst 4 respectively, which implies that the contribution of the persistent component from the non-thermal emission of the burst does not enhance during the peak of the burst. The flux from the non-thermal emission ($f_a$ method) varies between $\sim(4.4-7.6) \times 10^{-9}$ erg cm$^{-2}$ s$^{-1}$ for burst 3. The unabsorbed flux near the peak is $\sim4.2 \times 10^{-8}$ erg cm$^{-2}$ s$^{-1}$, which is $\sim$11 times higher than the persistent level. The non-burst persistent emission from the source is described with the disc blackbody model along with the thermally comptonized continuum model.
The non-burst persistent emission flux of the blackbody and comptonized component are $1.4 \times 10^{-9}$ and $2.9 \times 10^{-9}$ erg cm$^{-2}$ s$^{-1}$ respectively, during the pre-burst interval. 
 The upper limit on the source distance can be estimated using the peak flux during the burst. 

The unabsorbed blackbody flux  during the peak of the burst was $F_{max} = 3.4 \pm 0.15 \times 10^{-8}$ erg cm$^{-2}$ s$^{-1}$.
The source distance can be estimated using this value of Eddington luminosity using a linear equation of modified Stefan-Boltzmann law given by \citet{Le93}:
\begin{equation}
 L_{\text{Edd}} =  4\pi d^2 \xi F_{max}\\
\end{equation}
where $\xi$ is the anisotropy constant, which is taken as unity (assuming isotropic case), and $F_{max}$ is the unabsorbed peak flux. 

 During the peak of the burst, peak flux attained (4.2$\pm$0.1) $\times 10^{-8}$ erg cm$^{-2}$ s$^{-1}$ and (4.0$\pm$0.1) $\times 10^{-8}$ erg cm$^{-2}$ s$^{-1}$ for burst 3 and burst 4 respectively. The expansion of the photospheric radius expansion is not evident as seen from the evolution of the spectral parameters. Taking Eddington luminosity $\sim$3.8 $\times$10$^{38}$ erg  s$^{-1}$ \citep{Ku03}, the upper limit of the source distance is estimated to be $\sim$8.7 kpc.

We have also estimated the mass accretion rate at the surface of the neutron star. Depending on the local accretion rate, different thermonuclear ignition regimes can be determined. The accretion dominated luminosity assuming spherical accretion can be written as \citep{Ga06, Joh20}, 
\begin{equation}
L = 4\pi d^2 F \zeta = \frac{z \dot{M} c^2 }{(1+z)^3}  
\end{equation} 
where $\dot{M}$ is the mass accretion rate, $\zeta$ is the anisotropy factor, and $F$ is the pre-burst flux.  
We have calculated the mass accretion rate using a pre-burst flux of 4.3$\times$10$^{-9}$ erg cm$^{2}$ s $^{-1}$ and a source distance of $\sim$8.6 kpc.

\begin{equation}
\frac{(1+z)^3 4\pi d^2 F}{z \times c^2} \zeta = \dot{M}    
\end{equation}

\begin{equation}
\dot{M} = 6.6 \times 10^{-10} \times \frac{(1+z)^3 \times \zeta}{z} M_\odot yr^{-1}
\end{equation}
For an isotropic case, $\zeta=1$ and assuming $R=10$ km, $M = 1.4 M_\odot$, we get gravitational redshift on the surface of the NS is $(1+z) =\left(1-\frac{2GM}{c^2R}\right)^{-\frac{1}{2}}\simeq1.3$ and the mass accretion rate is estimated to be $\dot{M} \simeq4.87 \times$ 10$^{-9}$ $M_\odot$ yr$^{-1}$.

The change of burst properties of the individual system with the variation of accretion rates can be predicted by theoretical ignition models of hydrogen and helium-burning thermonuclear bursts.
For stable hot CNO cycle hydrogen burning, the mass accretion rate is related to the Eddington mass accretion rate as $\dot{M}$ $\ge$ 0.01 $\dot{M}_{Edd}$ \citep{Cu04, Ga06, Ga08}. In our study, we have found that $\dot{M}$ $\simeq$ 0.27 $\dot{M}_{Edd}$, which is higher than the estimated rates by \citet{Cu04} for stable CNO burning. At a lower accretion rate, $\dot{M}$ $\le$ 0.01 $\dot{M}_{Edd}$, the temperature of the burning layer is too low for stable hydrogen burning and the unstable hydrogen ignition leads to ``hydrogen-triggered'' bursts \citep{Cu04, Ga06}.

 The $\alpha$ factor for bursts defined from the ratio of persistent and burst fluence using the relation \citep{Ga08}:
\begin{equation}
    \alpha = \frac{F_{\text{persistent}}\times\Delta t_{r}}{E_{\text{burst}}}
\end{equation}
where $\Delta t_{r}$ is the recurrence time, $F_{\text{persistent}}$ is the bolometric flux of the persistent emission and $E_{\text{burst}}$ is the burst fluence. The burst fluence is estimated from the time-resolved spectra and the $F_{\text{persistent}}$ is obtained using the convolution model {\tt cflux} from the persistent spectra. One can categorize the sort of X-ray burst based on the theoretical value of $\alpha$. For a pure helium burst, $\alpha$ is nearly 100--150, whereas $\alpha$ is close to 40 for a hydrogen-rich X-ray burst, assuming each X-ray burst burns through all accreted matter \citep{Ga08}. Earlier, the $\alpha$ was also estimated to be $\sim45$ for MAXI J1816--195 using the Insight/HXMT and NICER observations \citep{Bu22, Che22}. The mass accretion rate, burst duration, recurrence time, and the $\alpha$ value indicate that the bursts are powered by a hydrogen-rich environment \citep{Le93}.

\section{Conclusion}
\label{con}
 We study the timing and spectral properties of the newly discovered millisecond pulsar MAXI J1816–195 during the 2022 outburst using simultaneous {\it NuSTAR} and {\it NICER} observations. Multiple Type-I thermonuclear bursts are detected using {\it NuSTAR} and {\it NICER} observations and the burst profiles are modelled using a sharp linear rise and exponential decay function. The duration of each burst was around $\sim$30 s, and a rise time of $\sim$5 s were found. The burst is found up to 20 keV energy range using {\it NuSTAR}. The burst profiles evolved significantly with energy, and a relatively long tail was observed in lower energy compared to higher energy.  The faster decay of the burst in a higher energy range implies that the temperature decrease as the burst evolves. The hardness ratio shows significant evolution during the thermonuclear burst. The {\it NICER} soft X-ray count rate (0.4--3 keV) is enhanced by $\sim$2.5 compared to the hard X-ray count rate (5--10 keV). Time-resolved spectroscopy is performed to study the evolution of different spectral parameters during the burst. The blackbody radius and flux increase towards the peak of the burst and show a trend to decrease in the decay phase of the burst. During the burst peak, the temperature and apparent emitting area of the blackbody reached a value of $\sim$2.1 keV and $\sim$12.5 km (for a source distance of $\sim$8.6 kpc) respectively. Assuming the empirical Eddington limit, the upper limit of the source distance for isotropic burst emission is estimated to be $\sim$8.7 kpc. The mass accretion rate is estimated to be $\simeq0.27 \dot{M}_{Edd}$, which implies stable burning of hydrogen in a hot CNO cycle.

\section*{Acknowledgements}
 We thank Dr. Tolga Guver for his useful suggestions, which help to improve the manuscript significantly. This research has made use of data obtained with {\it NuSTAR}, a project led by Caltech, funded by NASA, and managed by NASA/JPL, and has utilized the {\tt NUSTARDAS} software package, jointly developed by the ASDC (Italy) and Caltech (USA). This research has made use of the {\it MAXI} data provided by RIKEN, JAXA, and the {\it MAXI} team. We acknowledge the use of public data from the {\it NuSTAR}, {\it NICER}, and {\it Fermi} data archives. We thank the {\it NuSTAR} SOC Team for making this ToO observation possible. We are also thankful to the {\it NICER} team for the continuous monitoring of the source.

\section*{Data Availability}
The data used for this article are publicly available in the High Energy Astrophysics Science Archive Research Centre (HEASARC) at \\
\url{https://heasarc.gsfc.nasa.gov/db-perl/W3Browse/w3browse.pl}.



\bibliographystyle{mnras}




\appendix

\section{HR evolution}
Here, we have shown the evolution of the hardness ratio of MAXI J1816--195 during thermonuclear bursts using {\it NuSTAR}. 

\begin{figure*}
\centering{
\includegraphics[width=6.2cm, angle=270]{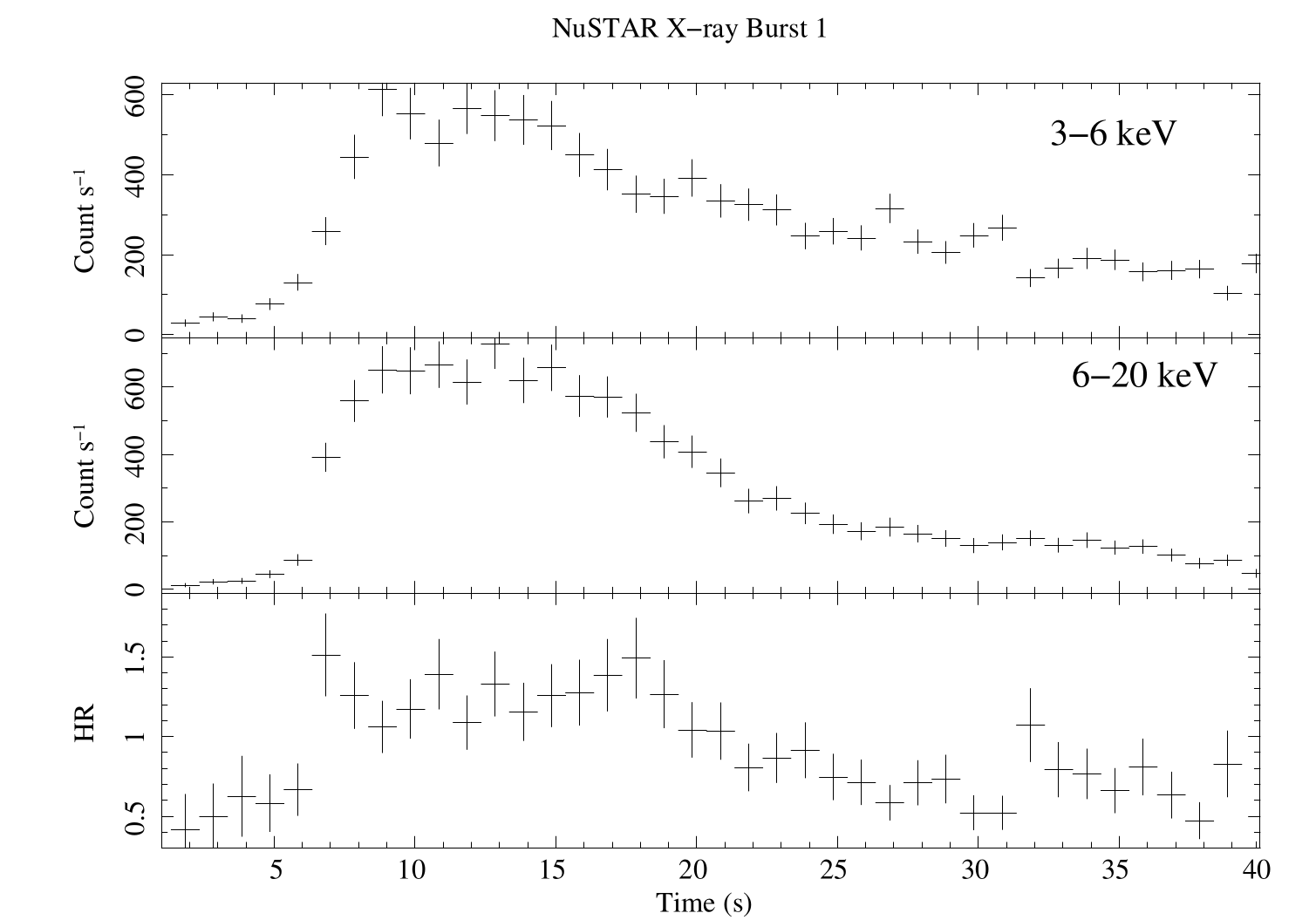}
\includegraphics[width=6.2cm, angle=270]{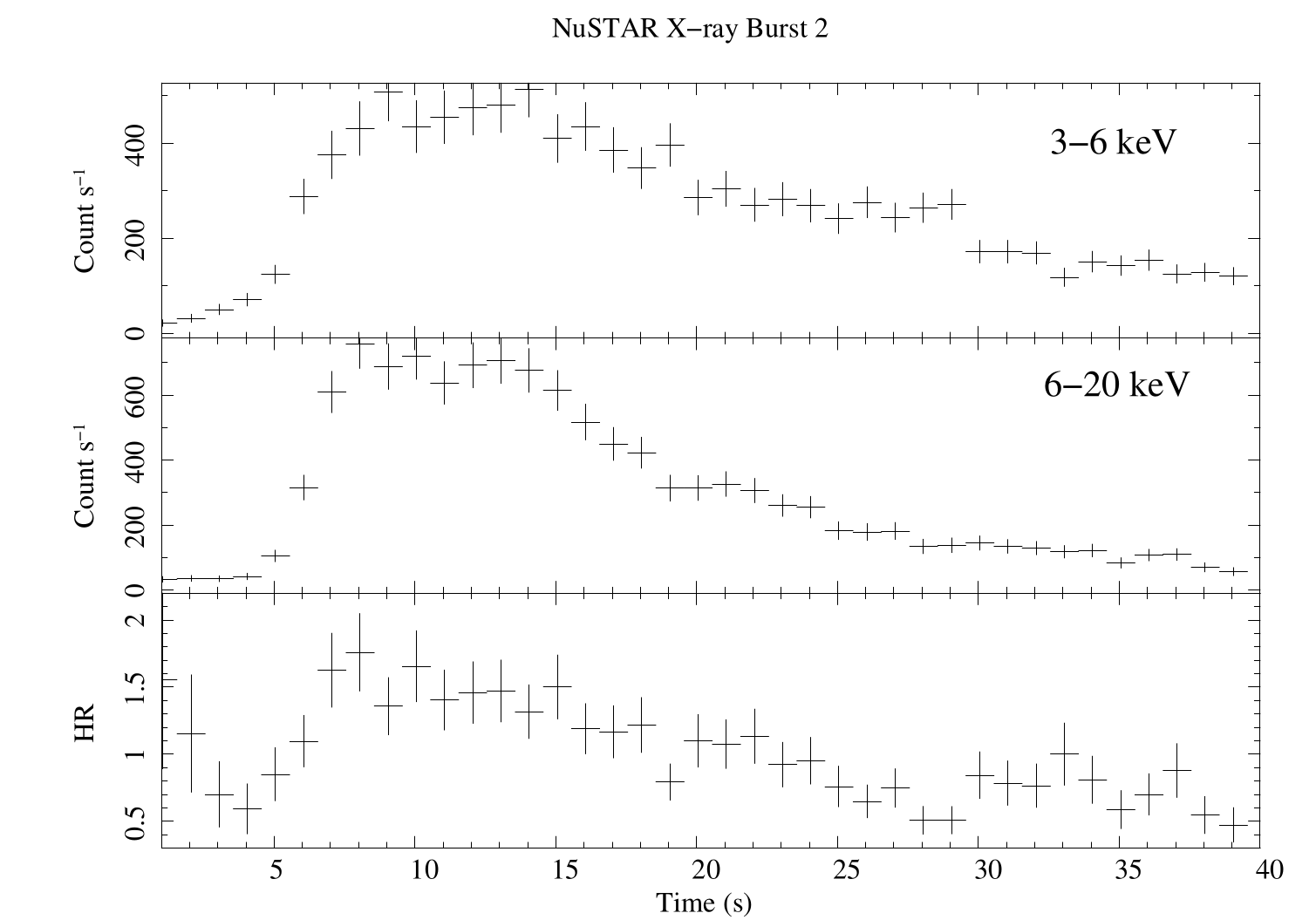}
\includegraphics[width=6.2cm, angle=270]{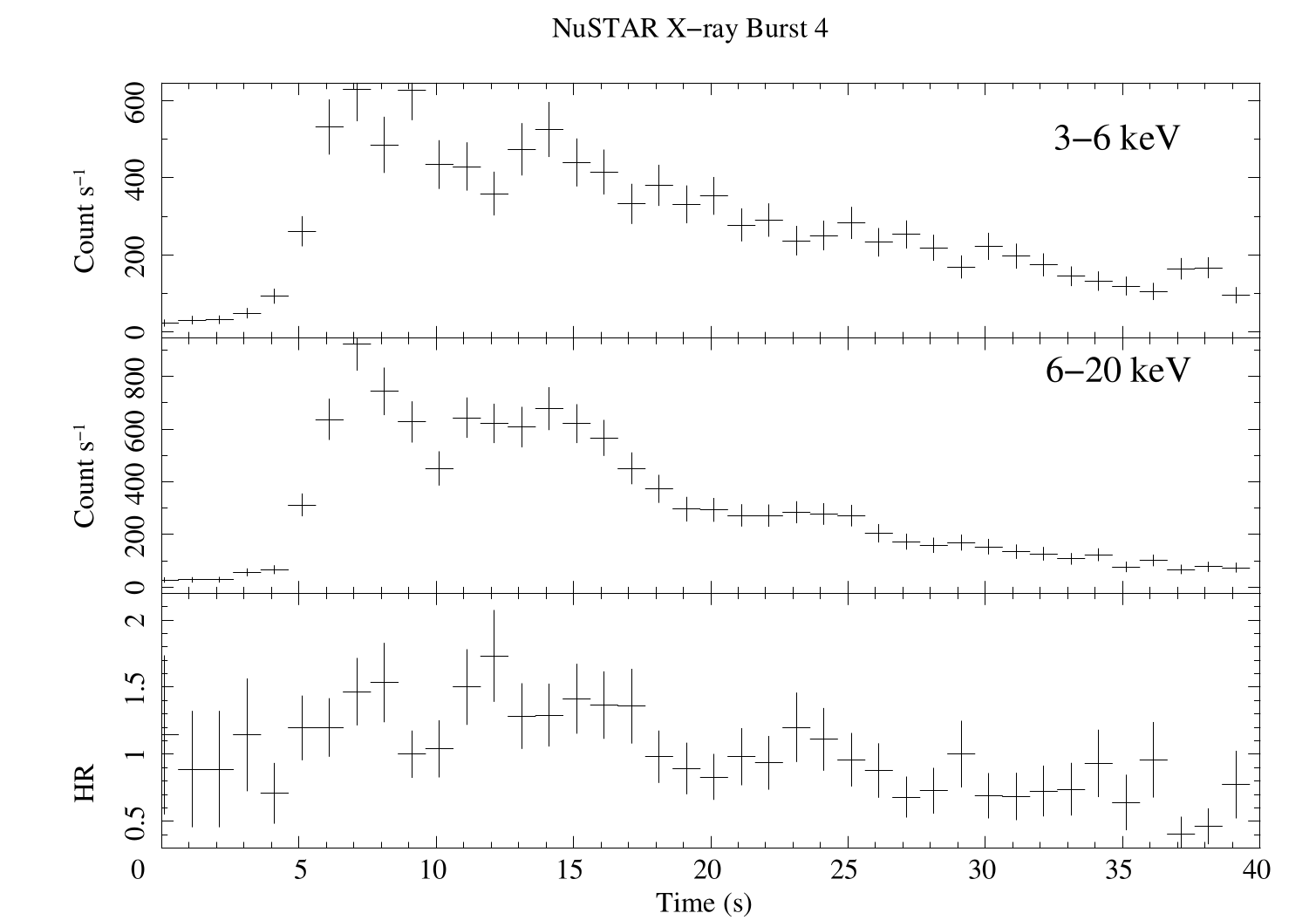}
\caption{{Evolution of hardness ratio of MAXI J1816--195 during thermonuclear bursts using {\it NuSTAR}. The first and second panels of each figure show the 1 s binned light curve in the energy bands of 3--6 keV and 6--20 keV, respectively. The bottom panel shows the variation of the hardness ratio (6--20 keV)/(3--6 keV) during bursts.}}
	 \label{fig:Hardness_evolution}}
\end{figure*}


\begin{table*}
\centering
\caption{Best-fitting parameters obtained through time-resolved analysis of {\it NICER} data for two bursts [{\tt XSPEC} model $\texttt{TBabs}\times \texttt{bbodyrad}$]. We model {\it NICER} spectra of 23 bins for burst-3 and burst-4 from MAXI J1816--195 with the variation of the persistent spectrum during the Burst using the $f_{a}$ method from \citet{Worpel2013}. The spectra from P2 to P18 are extracted for a time bin of 2 s and P19 to P24 are extracted for a time bin of 4 s. The hydrogen column density is fixed to 2.385$ \times 10^{22}$ cm$^{-2}$ during the fitting of time-resolved spectra.}
  
\begin{tabular}{|l|c|c|c|c|c|c|c|}
\hline
\hline
Burst phases & kT (keV) & Norm$_{kT}$ & Non-thermal emission ($f_a$)& $F_{Total}$ & $F_{bodyrad}$ &$F_{Non-thermal}$ &   $\chi^{2}$/dof \\
 
\hline
\hline
& & & Burst No. 3 & & & \\
\hline
\hline
P2 & 1.0$^{+0.20}_{-0.13}$ & 190$^{+170}_{-103}$ & 1.1$\pm0.2$ & ~7.0$\pm0.6$ & ~2.3$\pm 0.5$ & 4.7$\pm 0.5$ & 21.99/21 \\
P3 & 1.7$^{+0.17}_{-0.14}$ & 189$^{+55}_{-45}$ & 1.5$\pm 0.2$ & 21.6$\pm 0.8$ & 15.1$\pm1.0$ & 6.5$\pm0.6$ & 53.49/56 \\
P4 & 2.2$\pm0.13$ & 208$^{+35}_{-31}$ & 1.7$\pm0.2$ & 42.0$\pm1.3$ & 34.3$\pm1.5$ & 7.7$\pm0.7$ & 90.34/94 \\
P5 & 2.1$\pm0.13$ & 227$^{+39}_{-34}$ & 1.6$\pm0.2$ & 40.0$\pm 1.3$ & 32.7$\pm1.5$ & 7.0$\pm0.7$ & 96.85/90 \\
P6 & 2.0$\pm0.12$ & 261$^{+44}_{-38}$ & 1.6$\pm0.2$ & 40.4$\pm 1.2$ & 33.2$\pm1.4$ & 7.2$\pm0.7$ & 98.09/95 \\
P7 & 1.9$\pm0.12$ & 273$^{+46}_{-40}$ & 1.6$\pm 0.2$ & 39.0$\pm 1.2$ & 32.0$\pm1.4$ & 7.0$\pm0.7$ & 99.89/92 \\
P8 & 2.0$\pm0.12$ & 236$^{+41}_{-36}$ & 1.5$\pm0.2$ & 36.7$\pm 1.2$ & 30.0$\pm1.4$ & 6.8$\pm0.7$ & 85.51/87 \\
P9 & 1.9$\pm0.13$ & 233$^{+47}_{-40}$ & 1.4$\pm 0.2$ & 31.9$\pm 1.1$ & 25.8$\pm1.3$ & 6.1$\pm0.6$ & 81.95/77 \\
P10 & 1.9$^{+0.17}_{-0.14}$ & 192$^{+46}_{-39}$ & 1.7$\pm0.2$ & 30.3$\pm 1.1$ & 22.8$\pm1.4$ & 7.5$\pm0.6$ & 71.57/72 \\
P11 & 1.7$\pm0.12$ & 259$^{+57}_{-49}$ & 1.3$\pm 0.2$ & 24.3$\pm 0.9$ & 18.8$\pm1.0$ & 5.6$\pm0.6$ & 71.95/64 \\
P12 & 1.7$\pm0.14$ & 221$^{+59}_{-49}$ & 1.4$\pm 0.2$ & 23.2$\pm 0.9$ & 17.2$\pm1.1$ & 6.0$\pm0.6$ & 66.94/59 \\
P13 & 1.6$\pm0.14$ & 240$^{+71}_{-58}$ & 1.3$\pm 0.2$ & 20.4$\pm 0.8$ & 14.8$\pm1.0$ & 5.6$\pm0.6$ & 57.67/54 \\
P14 & 1.4$\pm0.11$ & 334$^{+100}_{-81}$ & 1.2$\pm 0.2$ & 17.6$\pm 0.7$ & 12.4$\pm0.8$ & 5.1$\pm0.6$ & 51.29/52 \\
P15 & 1.4$\pm 0.12$ & 290$^{+95}_{-76}$ & 1.2$\pm 0.2$ & 16.5$\pm 0.6$ & 11.2$\pm0.8$ & 5.3$\pm0.6$ & 52.68/48 \\
P16 & 1.5$^{+0.18}_{-0.14}$ & 191$^{+71}_{-55}$ & 1.4$\pm 0.2$ & 16.0$\pm 0.7$ & 10.0$\pm0.8$ & 6.0$\pm0.6$ & 41.77/44 \\
P17 & 1.4$^{+0.15}_{-0.12}$ & 240$^{+90}_{-69}$ & 1.1$\pm 0.2$ & 13.8$\pm 0.6$ & 8.9$\pm0.7$ & 4.9$\pm0.6$ & 44.28/39 \\
P18 & 1.5$^{+0.21}_{-0.15}$ & 176$^{+79}_{-59}$ & 1.2$\pm 0.2$ & 13.5$\pm 0.6$ & 8.3$\pm0.8$ & 5.2$\pm0.6$ & 42.19/37 \\
P19 & 1.3$\pm0.09$ & 246$^{+71}_{-58}$ & 1.1$\pm0.1$ & 11.2$\pm0.3$ & 6.5$\pm0.4$ & 4.7$\pm0.4$ & 66.79/68 \\
P20 & 1.2$\pm0.10$ & 231$^{+89}_{-69}$ & 1.0$\pm 0.1$ & ~9.0$\pm0.3$ & 4.6$\pm 0.3$ & 4.4$\pm 0.4$ & 59.51/55 \\
P21 & 1.1$\pm0.10$ & 271$^{+129}_{-96}$ & 1.0$\pm 0.2$ & ~8.1$\pm0.3$ & 3.5$\pm0.3$ & 4.6$\pm0.4$ & 48.21/51 \\
\hline
\hline
& & & Burst No. 4 & & & \\
\hline
\hline
P2 & 2.1$\pm 0.15$ & 190$^{+39}_{-34}$ & 1.6$\pm0.2$ & 34.0$\pm 1.2$ & 27.0$\pm 1.4$ & 7.0$\pm0.6$ & 79.66/79 \\
P3 & 2.1$\pm 0.15$ & 190$^{+39}_{-34}$ & 1.6$\pm0.2$ & 34.0$\pm 1.2$ & 27.0$\pm 1.4$ & 7.0$\pm0.6$ & 80.66/79 \\
P4 & 2.1$\pm 0.13$ & 218$^{+38}_{-33}$ & 1.7$\pm0.2$ & 40.1$\pm 1.3$ & 32.5$\pm 1.5$ & 7.6$\pm 0.7$ & 86.29/92 \\
P5 & 2.1$\pm 0.13$ & 227$^{+40}_{-35}$ & 1.7$\pm0.2$ & 40.3$\pm 1.3$ & 33.0$\pm1.5$ & 7.4$\pm0.7$ & 93.54/91 \\
P6 & 2.1$\pm 0.13$ & 213$^{+37}_{-32}$ & 1.7$\pm0.2$ & 38.9$\pm 1.2$ & 31.5$\pm1.4$ & 7.3$\pm0.7$ & 88.46/89 \\
P7 & 2.0$\pm 0.13$ & 233$^{+44}_{-39}$ & 1.6$\pm0.2$ & 36.8$\pm 1.2$ & 29.6$\pm1.5$ & 7.2$\pm0.7$ & 89.22/86 \\
P8 & 2.0$\pm 0.13$ & 231$^{+47}_{-40}$ & 1.6$\pm0.2$ & 34.8$\pm 1.2$ & 27.6$\pm1.4$ & 7.2$\pm0.6$ & 88.68/82 \\
P9 & 1.7$\pm 0.10$ & 330$^{+63}_{-55}$ & 1.1$\pm0.2$ & 27.2$\pm 0.9$ & 22.5$\pm1.1$ & 4.7$\pm0.6$ & 75.37/73 \\
P10 & 1.8$\pm 0.14$ & 221$^{+55}_{-46}$ & 1.5$\pm0.2$ & 25.9$\pm 0.9$ & 19.2$\pm1.2$ & 6.8$\pm0.6$ & 67.71/66 \\
P11 & 1.7$\pm 0.12$ & 238$^{+54}_{-46}$ & 1.2$\pm 0.2$ & 23.0$\pm 0.9$ & 18.0$\pm1.0$ & 5.1$\pm0.6$ & 59.56/60 \\
P12 & 1.6$^{+0.15}_{-0.12}$ & 246$^{+69}_{-57}$ & 1.3$\pm 0.2$ & 21.0$\pm 0.8$ & 15.5$\pm1.0$ & 5.5$\pm0.6$ & 61.87/56 \\
P13 & 1.5$\pm0.12$ & 262$^{+73}_{-60}$ & 1.1$\pm 0.2$ & 18.2$\pm 0.7$ & 13.4$\pm0.9$ & 4.9$\pm0.6$ & 53.64/51 \\
P14 & 1.6$\pm 0.13$ & 203$^{+59}_{-48}$ & 1.4$\pm 0.2$ & 19.1$\pm 0.7$ & 12.7$\pm0.9$ & 6.3$\pm0.6$ & 57.10/51 \\
P15 & 1.5$^{+0.15}_{-0.12}$ & 232$^{+77}_{-61}$ & 1.3$\pm0.2$ & 16.8$\pm 0.7$ & 11.0$\pm0.8$ & 5.7$\pm0.6$ & 48.50/46 \\
P16 & 1.3$\pm 0.1$ & 362$^{+112}_{-90}$ & 0.9$\pm 0.2$ & 14.1$\pm 0.6$ & 10.2$\pm0.6$ & 3.9$\pm0.6$ & 47.64/43 \\
P17 & 1.3$\pm 0.1$ & 292$^{+112}_{-86}$ & 1.0$\pm 0.2$ & 12.5$\pm 0.5$ & 8.0$\pm0.6$ & 4.5$\pm0.6$ & 40.06/38 \\
P18 & 1.3$^{+0.15}_{-0.12}$ & 240$^{+100}_{-76}$ & 1.1$\pm 0.2$ & 12.0$\pm0.5$ & 7.2$\pm0.5$ & 4.8$\pm0.6$ & 40.14/35 \\
P19 & 1.2$\pm 0.1$ & 285$^{+79}_{-65}$ & 0.9$\pm 0.1$ & 10.4$\pm0.3$ & 6.3$\pm0.3$ & 4.1$\pm0.4$ & 65.57/64 \\
P20 & 1.2$^{+0.10}_{-0.05}$ & 251$^{+84}_{-67}$ & 0.9$\pm 0.1$ & ~9.2$\pm0.3$ & 5.1$\pm0.3$ & 4.1$\pm0.4$ & 62.93/57 \\
P21 & 1.0$\pm 0.08$ & 293$^{+118}_{-91}$ & 1.0$\pm 0.2$ & ~7.7$\pm0.2$ & 3.5$\pm0.2$ & 4.2$\pm0.4$ & 53.25/49 \\
P22 & 1.0$\pm 0.10$ & 268$^{+136}_{-99}$ & 0.9$\pm 0.2$ & ~6.6$\pm0.2$ & 2.6$\pm0.2$ & 4.0$\pm0.4$ & 45.39/42 \\
P23 & 0.9$\pm 0.09$ & 291$^{+160}_{-116}$ & 0.9$\pm 0.2$ & ~6.1$\pm0.2$ & 2.2$\pm 0.2$ & 4.0$\pm0.4$ & 44.49/40 \\
P24 & 0.8$^{+0.12}_{-0.08}$ & 355$^{+259}_{-177}$ & 0.9$\pm 0.2$ & ~5.6$\pm0.2$ & 1.8$\pm 0.2$ & 3.8$\pm0.4$ & 39.28/35 \\
\hline
\hline
\label{tab:tab4}
\end{tabular}\\
{\bf *} : All the errors are 90\% significant and calculated using the MCMC approach in {\tt XSPEC}\\
{\bf **}: All the flux values are unabsorbed and calculated for the energy band 0.1--10 keV and in the unit of 10$^{-9}$ ergs cm$^{-2}$ s$^{-1}$

\end{table*}
\pagestyle{plain}

\bsp	
\label{lastpage}
\end{document}